\documentclass[9.5pt,letter,twoside]{rho-class/rho}
\usepackage[english]{babel}

\usepackage{tikz}
\usepackage{graphicx}
\usepackage{float}
\usepackage{natbib}
\usepackage{pgfplots}
\pgfplotsset{compat=1.17}
\usepackage{tabularx}
\usepackage{multirow}
\usepackage{placeins}

\newcommand{\order}[1]{\textit{O}(#1)}

\makeatletter
\newcommand{\dropcapinitial}[1]{%
    \let\firstletter\@empty
    \renewcommand{\firstletter}{\expandafter\@car#1\@nil}%
    \lettrine[lines=2,lraise=0,findent=2pt, nindent=0em]{{\dropcapfont{\firstletter}}}{\expandafter\@gobble#1}
}
\makeatother

\newenvironment{axiomlist}
    {\begin{list}{}{\setlength{\leftmargin}{0pt}\setlength{\itemsep}{6pt}\setlength{\parsep}{3pt}}}
    {\end{list}}

\DeclareRobustCommand{\nec}{\ensuremath{\mathrm{N}}}
\DeclareRobustCommand{\Nc}{\ensuremath{\mathrm{N}_{\mathrm{c}}}}
\DeclareRobustCommand{\poss}{\ensuremath{\Pi}}
\DeclareRobustCommand{\ign}{\ensuremath{H_{\Pi}}}
\DeclareRobustCommand{\pinorm}{\ensuremath{\pi'}}

\newcommand{\spc}[1]{\textsc{#1}}   

\colorthemePossVerifPurple
\colorthemePossVerifGreen

\usepackage{tcolorbox}
\newtcolorbox{workedexample}[1][]{%
    colback=rationalecolor!8,
    colframe=rationalecolor!50!black,
    coltitle=white,
    fonttitle=\bfseries\sffamily,
    title={#1},
    sharp corners,
    boxrule=0.5pt,
    left=4pt, right=4pt, top=2pt, bottom=2pt
}

\setbool{rho-abstract}{true}
\setbool{corres-info}{true}

\journalname{Preprint (submission in preparation)}
\title{Possible, Yes; Ignorant, Perhaps: A Scorecard for Possibilistic Forecasts}

\author[1,2]{John R.\ Lawson}

\affil[1]{Bingham Research Center, Utah State University, Vernal, Utah, United States of America}
\affil[2]{Department of Mathematics and Statistics, Utah State University, Logan, Utah, United States of America}

\dates{Draft date: 2 April 2026}

\leadauthor{John R.\ Lawson}
\footinfo{Possibilistic forecast verification}
\smalltitle{\LaTeX\ Preprint}
\institution{Utah State University}
\theday{2 April 2026\newline{}Submitted to J.\ Atmos.\ Sci.}

\corres{Correspondence to John R.\ Lawson}
\email{john.lawson@usu.edu}

\license{Rho LaTeX Class \ccLogo\ This document is licensed under Creative Commons CC BY 4.0. Template further modified by the author.}

\begin{abstract}
Probabilistic forecasts must sum to unity and cannot
express ``I don't know.''
Possibility theory relaxes this constraint:
a subnormal distribution explicitly measures how much of
the plausibility budget remains unassigned, ignorance
signal that probability cannot represent.
This paper develops a verification framework for such
forecasts, centred on a five-number scorecard that
separately diagnoses whether the forecast pointed at
the right outcome (depth-of-truth), how sharply
(diffuseness, support margin), how confidently
(ignorance), and how dominantly (conditional necessity).
A possibility-to-probability conversion preserves
ignorance for familiar frequency-based scoring;
categorical threshold scores (POD, FAR, CSI, etc.)
connect to operational practice.
Together, these three complementary
facets---possibilistic, probabilistic, and
categorical---expose failure modes invisible to any
single metric.
Storm Prediction Center convective outlook categories
serve as the running example throughout;
a synthetic reforecast demonstrates diagnostic
visualisations and scorecard interpretation.
Ignorance is better expressed than repressed.
\end{abstract}


\begin{document}
    \maketitle
    \thispagestyle{firststyle}
    \setcounter{tocdepth}{2}
    \tableofcontents



\section*{Significance Statement}
\begingroup
\hyphenpenalty=10000
\exhyphenpenalty=10000
When a forecasting system is uncertain, that uncertainty
is decision-relevant information.
Yet probabilistic forecasts, which must sum to unity,
cannot distinguish a confident prediction from an
ignorant one.
Possibility theory fills this gap: a subnormal
distribution explicitly measures how uncertain the
system is about its own guidance.
This paper introduces the first verification framework
for possibilistic forecasts, centred on a five-number
scorecard that diagnoses whether the forecast pointed
at the right outcome, how sharply, and how honestly.
A bridge to probabilities and familiar categorical
metrics (POD, FAR, CSI) connects the framework to
existing practice.
The result: forecast uncertainty is diagnosed rather
than hidden, enabling targeted system improvement and
more transparent risk communication.
\endgroup


\section{Introduction}\label{sec:introduction}

Consider two forecasts, each assigning equal odds. A statistician
estimates a fair coin at 50:50 after one hundred flips---confident,
grounded in data. A media pundit predicts 50:50 for a
sports match played between two obscure teams that have
never met---uncertain, grounded
in ignorance. Probability theory assigns both the same distribution;
possibility theory does not. The coin forecast is \textit{normal},
fully endorsing its estimate; the match forecast is
\textit{subnormal}, marking that equal odds arose from absence of
evidence rather than weight of evidence.
The possibility $\Pi$,
an effective upper-bound on uncertain probabilities
\citep{Dubois1992-gd},
is most beneficial to the most risk-averse end-users,
as high value is place on the caveat the worst-case
scenario could be even worse due to uncertainty
(that are hidden by sharp, traditional probabilities).

Hazard forecasting increasingly relies on probabilistic methods to
communicate uncertainty: calibrated probability
distributions support principled decision-making and allow rigorous
verification through proper scoring rules
\citep{Hendrickson1971-lx,Gneiting2007-ob}.
Yet probability demands
precise quantification, where each outcome receives a single number
representing relative frequency (or belief), and those numbers
must sum to unity. Not all forecasting systems produce output that
satisfies these constraints. This paper uses Storm Prediction Center
(SPC) convective outlook categories as a running example. SPC
outlooks assign one of five risk categories---\spc{MRGL} (Marginal),
\spc{SLGT} (Slight), \spc{ENH} (Enhanced), \spc{MDT} (Moderate), and
\spc{HIGH}---or \spc{NONE} if no categorical risk is assigned
(Section~\ref{sec:poss_primer} gives full definitions). The scenarios
below are thought experiments framed as a
"shadow prediction" of Day 1 SPC Outlooks that then
acted as "verification" to match like-for-like
for pedagogy's sake.
Possibility theory \citep{Dubois1988-nh}
provides a mathematical framework for graded
compatibility of outcomes with available evidence
without imposed additivity requirements.
This paper presents a scorecard that capture many
orthogonal aspects of forecast quality in pursuit of
optimisation and evaluation of forecasts under high
uncertainty.

\subsection{Possibility theory and hazard verification}

Possibility theory extends probability
\citep{Zadeh1978-je}:
for any event $A$, the interval
from necessity $\nec(A)$ to possibility $\poss(A)$
brackets the range of consistent probabilities
(Section~\ref{sec:poss_axioms};
Table~\ref{tab:notation} collects all symbols).
Possibility distributions arise naturally in several
operational settings;
for example, expert-rule systems and fuzzy inference
engines \citep[e.g.,][]{Chevrie1998-go} produce
membership functions that are possibility distributions
by construction \citep{Klir1995-va}.
Possibility functions can also be assigned more
subjectively with more reliance on human confidence
than calculated support in data.
In severe-weather forecasting, a
forecaster may wish to express that \spc{MDT} risk is
plausible but not certain,
while acknowledging that the mesoscale environment is poorly
sampled---a statement that maps cleanly onto a subnormal possibility
distribution $\pi$ but awkwardly onto probabilities. Assigning
$\pi(\spc{MDT}) = 0.7$ says ``\spc{MDT} is fairly compatible with
available evidence''; it does not say ``the long-run frequency of
\spc{MDT} events is 70\%.''

For subnormal distributions ($\max_\omega \pi(\omega) < 1$),
possibilistic ignorance $\ign = 1 - \max(\pi)$ quantifies the gap
between the system's best-supported outcome and full endorsement---a
signal that probability, which must sum to unity, cannot represent.
Normalising a subnormal distribution to
$\pi' \equiv \max(\pi) = 1$ erases this
signal as it conflates ``the
system considers \spc{ENH} most likely'' with ``the system is
confident'' (Section~\ref{sec:naive_normalisation} develops this
contrast).

\subsection{Background}

Possibility theory originates with \citet{Zadeh1978-je},
who interpreted fuzzy membership functions as
possibility distributions. It was formalised
into a full uncertainty calculus by \citet{Dubois1988-nh}. The
bracketing $\nec(A) \leq P(A) \leq \poss(A)$ is well
suited applied to risk analysis under incomplete
information.
\citet{Klir1995-va} couches possibility within a
broader group of information theories,
clarifying some relationship between possibilistic
and probabilistic uncertainty measures.

The author underscores that the choice of objectively
optimal evaluation rules is subjective.
Ultimately, metrics must be chosen to test the
narrative of forecast quality.
Verification of probabilistic forecasts often uses
\textit{proper} scoring rules,
where forecasters cannot hedge
(e.g., incorrectly adjusting forecasts to reduce
peaks but accumulate penalty).
\citet{Roulston2002-eq} introduced ignorance,
or relative entropy \citep{Cover2012-di} and logarithmic score
\citep{Good1952-st}, measuring forecast value as
the reduction in Shannon entropy relative to a
climatological baseline \citep{Peirolo2011-sl}.
The preceding papers formalise decomposition of overall skill into
uncertainty, discrimination, and reliability components
analogous to Brier-type decomposition
\citep[see also][]{Roulston2002-eq,Benedetti2010-sa,Weijs2010-hg, Todter2012-ou}).
These tools are designed for probability vectors,
not possibility distributions that violate the
additivity axiom.
The broader meteorological verification literature
\citep{Murphy1984-ox,Jolliffe2003-xs,Wilks2011-vw}
provides the conceptual vocabulary
(reliability, resolution, discrimination)
adapted below.
Second-order uncertainty is not a new concept:
adjacent frameworks and concepts,
such as Dempster--Shafer evidence theory
\citep{Yager2005-at}, the pignistic transformation
\citep{Smets1990-nf}, and imprecise probabilities
\citep{Walley1982-pm}, share the intuition that a single
probability may not capture epistemic uncertainty
(what a forecaster actually knows and doesn't know;
Section~\ref{sec:three_component}).

Nonetheless, there is motivation for encouraging
explicit preservation of forecast uncertainty when
communicating risk,
and more so for risk-averse stakeholders that benefit
most from upper-bounds (possibilities) of the
undesired scenario.
Existing approaches that normalise the distribution
to match axiomata of possibility theory erases the
ignorance signal.
Further, possibility theory is unfamiliar in many
applied fields,
and clear visualisations and necessary to communicate
a larger set of forecast metrics.
The current paper explains a framework with three
components: a
five-number scorecard for possibilistic evaluation, a
possibility-to-probability conversion that preserves ignorance as an
explicit outcome, and a categorical verification facet using
threshold-based scores familiar to operational meteorologists.
Traditional categorical verification
(POD, FAR, CSI) discussed in \citet{Green1966-xl} and \citet{Jolliffe2003-xs} trades most
distributional information for low-complexity
interpret ion;
by reducing a possibility distribution to its peak
category;
Section~\ref{sec:tripartite_value} formalises this
observation and discusses shortcomings in each
verification approach.
The remainder develops the theory modification
(Sections~\ref{sec:poss_primer}--\ref{sec:native_verification}),
demonstrates three complementary verification facets
(Section~\ref{sec:tripartite_value}), walks through worked examples
(Section~\ref{sec:worked_examples}),
and concludes with applications and future directions
(Section~\ref{sec:discussion}).


\section{Possibility Theory Foundations}\label{sec:poss_primer}

This section introduces possibility theory for
atmospheric scientists, and no prior knowledge of fuzzy logic or
uncertain probability is assumed.
We employ a running example of a "shadow"
possibilistic convective outlook that assigns
possibility values to the five SPC categories
(plus \spc{NONE}) introduced in
Section~\ref{sec:introduction} to illustrate
definition and aid interpretation.
The categories are mutually exclusive:
given one point in time and space,
exactly one category describes the observed
outcome.
In all examples below,
the outlook at a single point serves as
the observation.

\subsection{Axioms and Definitions}\label{sec:poss_axioms}

Let $\Omega$ denote the
\textit{universe of discourse},
or the complete set of mutually exclusive and
collectively exhaustive outcomes for a
variable of interest
(e.g., SPC convective outlook categories
$\Omega = \{\spc{NONE},\, \spc{MRGL},\, \spc{SLGT},\, \spc{ENH},\,
\spc{MDT},\, \spc{HIGH}\}$). A \textit{possibility distribution}
$\pi: \Omega \to [0,1]$\allowbreak{} assigns to each
outcome $\omega \in \Omega$\allowbreak{} a degree of
possibility $\pi(\omega)$ or \textit{truthiness}
value that satisfies the following.

\begin{axiomlist}
\item \textbf{Something must be possible.} There exists at least one
    outcome $\omega^* \in \Omega$ such that $\pi(\omega^*) > 0$.
\item \textbf{Normalisation.} If
    $\max_{\omega \in \Omega} \pi(\omega) = 1$, the distribution is
    \textit{normal}; otherwise it is \textit{subnormal}. Classical
    possibility theory \citep{Dubois1988-nh}
    requires normality:
    at least one outcome to be fully possible
    ($\max \pi = 1$),
    grounding necessity in complete commitment.
    This paper relaxes that requirement --- a
    subjective modelling choice permitted by
    original formulation \citep{Zadeh1978-je}
    and adopted to forecasting systems that
    often encounter conditions outside their
    training domain. Subnormality encodes
    assessment derived from support in data,
    with the remaining plausibility budget
    ($\ign$) explicitly flagging the
    admitted-unsure portion
    (Section~\ref{sec:subnormal}).
\item \textbf{Max-additivity:} For any events $A, B \subseteq \Omega$
    with $A \cap B = \emptyset$,
    $\Pi(A \cup B) = \max\bigl(\Pi(A),\, \Pi(B)\bigr)$.
    This axiom distinguishes possibility from probability: possibility
    is bounded by the maximum rather than the sum of its components.
\end{axiomlist}

From possibility distribution $\pi$ that may or
may not be normalised ($\pi = \pi'$),
dual measures are defined for event
$A \subseteq \Omega$:

\begin{align}
\Pi(A) &= \max_{\omega \in A} \pi(\omega)
    \label{eq:poss} \\
N(A) &= 1 - \Pi(\neg A)
     = 1 - \max_{\omega \notin A} \pi(\omega)
    \label{eq:nec}
\end{align}

\textbf{Interpretation:}
\begin{itemize}[leftmargin=2em]
\item $\Pi(A)=1$: Event $A$ is entirely
    plausible given current knowledge.
\item $\Pi(A)=0$: Event $A$ is impossible (inconsistent with available
    evidence).
\item $N(A)=1$: Event $A$ is certain (all alternatives are ruled out).
\item $N(A)=0$: Event $A$ is not certain (at least one alternative
    remains plausible).
\end{itemize}

For normal distributions ($\max\pi=1$), the interval
$[N(A),\, \Pi(A)]$ brackets all compatible probabilities
\citep{Dubois1988-nh,Dubois2007-dd}.
If $\pi \neq \pi'$ (subnormality),
this coherence can be violated,
hence Section~\ref{sec:three_component}
introduces conditional necessity $\Nc$ to restore
a metric of certainty operating on the
normalised shape.
For reference,
Table~\ref{tab:notation} collects recurring
symbols.

\begin{table}[htbp]
\centering
\caption{Notation reference for possibilistic verification. ``Form''
    indicates whether the quantity uses the raw ($\pi$) or normalised
    ($\pinorm$) distribution.}
\label{tab:notation}
\small
\begin{tabularx}{\columnwidth}{lclX}
\toprule
Symbol & Form & Eq. & Meaning \\
\midrule
$\Omega$ & --- & --- & Universe of discourse \\
$\pi$ & raw & --- & Raw (possibly subnormal) distribution \\
$\pinorm$ & norm & --- & Normalised distribution: $\pi / m$ \\
$m$ & raw & --- & Commitment (peak possibility): $\max_{\omega} \pi(\omega)$ \\
$\poss(A)$ & raw & \ref{eq:poss} & Possibility of event $A$ \\
$\nec(A)$ & raw & \ref{eq:nec} & Necessity of event $A$ \\
$\Nc(A)$ & norm & \ref{eq:cond_nec} & Conditional necessity of $A$ \\
$\ign$ & raw & \ref{eq:ignorance} & Possibilistic ignorance (not Shannon entropy): $1 - m$ \\
$\alpha^*$ & norm & \ref{eq:alpha_star} & Depth-of-truth: $\pinorm(c_{\mathrm{obs}})$ \\
$\eta$ & norm & \ref{eq:ns} & Diffuseness: $(1/K)\sum_c \pinorm(c)$ \\
$\delta$ & norm & \ref{eq:rg} & Support margin: $\alpha^* - \eta$ \\
$\Nc^*$ & norm & \ref{eq:nc_star} & Cond.\ necessity of truth \\
\bottomrule
\end{tabularx}
\end{table}

\subsection{Subnormal Distributions and Ignorance}\label{sec:subnormal}

In operational forecasting,
subnormality (awareness of `known unknowns')
arises when a system encounters conditions
outside its training domain or if presented
with conflicting evidence and weak support
for all outcomes.
Rather than hiding this deficiency behind
enforced normalisation,
preserving a measure of ignorance makes
explicit the system's knowledge
gap \citep{Oussalah2002-os}.

Define the (possibilistic) \textit{ignorance} measure \ign{} as the
residual possibility ``mass'' not covered in $\pi$:
\begin{equation}
H_\Pi = 1 - \max_{\omega \in \Omega} \pi(\omega) \label{eq:ignorance}
\end{equation}

\textbf{Interpretation:} \ign{} quantifies the fraction of plausibility
unaccounted for by the model's knowledge. High ignorance
($H_\Pi \approx 1$) signals that the forecasting system encounters
conditions outside its coverage or that conflicting evidence produces
weak support for all outcomes. Low ignorance ($H_\Pi \approx 0$)
indicates the system assigns high possibility to at least one
outcome, and an impossible event can be ruled
out due to known plausibility of a different outcome, rather than
spurious confidence after normalisation.

\begin{figure}[t]
\centering
\includegraphics[width=\columnwidth]{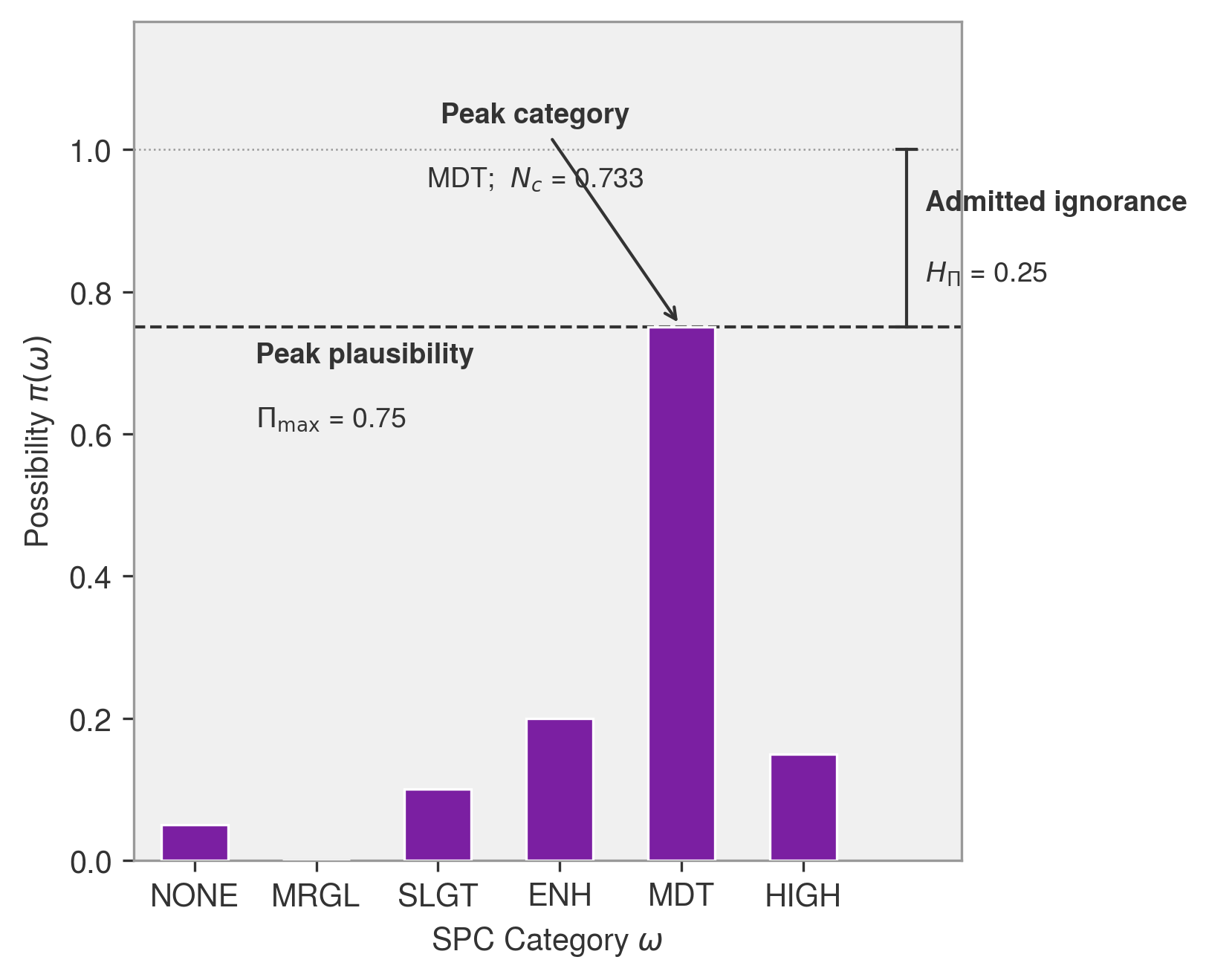}
\caption{Anatomy of a subnormal possibility distribution over SPC
    categories. Purple bars show raw possibility $\pi(\omega)$. The
    dashed line marks $\Pi_{\max} = \max(\pi)$; the gap from
    $\Pi_{\max}$ to $1.0$ is the ignorance $\ign$
    (Eq.~\ref{eq:ignorance}). Conditional necessity \Nc{} for the peak
    category measures how strongly that category dominates the
    runner-up after normalisation (Eq.~\ref{eq:cond_nec}); the
    annotated value $0.733$ reflects $1 - 0.20/0.75$, where $0.20$
    is the runner-up (\spc{ENH}) and $0.75$ is the peak (\spc{MDT}).}
\label{fig:poss_anatomy}
\end{figure}

Figure~\ref{fig:poss_anatomy} illustrates a subnormal distribution
annotated with $\Pi_{\max}$, \ign{}, and \Nc{} for the peak category.

Forcing a probability
distribution to sum to unity masks the distinction between ``the
system predicts \spc{MDT}'' and ``the system has insufficient
information to rule out any category.'' Preserving subnormality trades
formal completeness for operational transparency: the forecaster sees
not only which categories are plausible but how much confidence the
system places in the entire guidance.

\textbf{Example 1 (sharp \spc{MDT} forecast):}
Consider a classic severe-weather setup in which all ingredients for
significant severe thunderstorms are present and co-located.
The resulting possibility distribution is:
\[
\pi(\omega) = \begin{cases}
0.05 & \omega = \spc{NONE} \\
0.00 & \omega = \spc{MRGL} \\
0.10 & \omega = \spc{SLGT} \\
0.20 & \omega = \spc{ENH}  \\
0.75 & \omega = \spc{MDT}  \\
0.15 & \omega = \spc{HIGH}
\end{cases}
\]
with $\ign = 1 - 0.75 = 0.25$ and
$\Nc(\spc{MDT}) = 1 - 0.20/0.75 = 1 - 0.267 = 0.733$, where $0.20$
is the runner-up raw possibility (\spc{ENH}). Summarised, \spc{MDT}
is the most plausible outcome ($\Pi = 0.75$) with moderate ignorance
($\ign = 0.25$), while conditional necessity
of $0.73$ indicates that,
amongst scenarios the system covers,
\spc{MDT} is the dominant category.

\textbf{Example 2 (ambiguous \spc{MRGL}/\spc{SLGT} forecast):}
Now consider a borderline severe-weather setup where instability is
moderate, shear is variable, and convective initiation is uncertain.
The resulting possibility distribution is:
\[
\pi(\omega) = \begin{cases}
0.30 & \omega = \spc{NONE} \\
0.50 & \omega = \spc{MRGL} \\
0.50 & \omega = \spc{SLGT} \\
0.00 & \omega = \spc{ENH}  \\
0.00 & \omega = \spc{MDT}  \\
0.00 & \omega = \spc{HIGH}
\end{cases}
\]
with $\ign = 1 - 0.5 = 0.5$ and
$\Nc(\spc{MRGL}) = \Nc(\spc{SLGT}) = 0.0$. Interpretation:
\spc{MRGL} and \spc{SLGT} are equally plausible ($\Pi = 0.5$ each),
high ignorance ($\ign = 0.5$) indicates that the system struggles with
this meteorological scenario, and neither outcome has conditional
necessity.

Blind normalisation in Example~2 would inflate $\pi(\spc{MRGL})$ and
$\pi(\spc{SLGT})$ to 1.0, hiding the fact that the forecasting system
struggles with this meteorological scenario. Preserving subnormality
makes these knowledge gaps explicit.

For clarity, hereon distributions are written as ordered tuples
$(\spc{NONE},\, \spc{MRGL},\, \spc{SLGT},\, \spc{ENH},\, \spc{MDT},\,
\spc{HIGH})$ when context is clear:\ e.g., Example~1 is
$\pi = (0.05,\, 0,\, 0.1,\, 0.2,\, 0.75,\, 0.15)$.

\subsection{Three-Component Uncertainty}\label{sec:three_component}

An analogue of ``classical'' necessity (requiring a normalised
possibility distribution $\pinorm$) supplements \poss{} and \ign{} for
each SPC category $A$ by computing necessity after normalisation of
known outcomes. The three components are:

\begin{enumerate}[leftmargin=2em]
\item \textbf{Possibility, $\poss(A)$, $\pi$:} raw (unnormalised)
     evidence-based plausibility of event $A$
     from Equation~\eqref{eq:poss}.
\item \textbf{Ignorance, \ign{}:} global ignorance measure in
    Equation~\eqref{eq:ignorance},
    indicating inherent deficiency of the
    forecasting system in capturing the
    notional ``true'' possibility
    distribution: an honest measurement of a `known unknown'.
\item \textbf{Conditional Necessity, $\Nc(A)$:} Certainty of
    event $A$ computed \textit{after} normalising the distribution of
    known outcomes:
\begin{equation}
N_c(A) = 1 - \frac{\max_{\omega \notin A} \pi(\omega)}%
    {\max_{\omega \in \Omega} \pi(\omega)} \label{eq:cond_nec}
\end{equation}
This measures certainty conditional on the system's covered scenarios.
Equivalently, $\Nc(A) = 1 - \max_{\omega \notin A} \pinorm(\omega)$,
complement of the largest normalised
possibility among alternatives to~$A$.
\end{enumerate}

All three quantities derive from the raw
possibility distribution $\pi$:
\poss{} is the peak value for each category,
\ign{} measures the global gap to normality,
and \Nc{} is computed after renormalisation.
Though not statistically independent,
they decompose the forecast signal into
complementary facets of decision-relevant
information.
These three components serve the forecaster
\textit{before} an outcome
is known: for any hypothetical event $A$, the triple
$(\Pi(A),\; \ign,\; \Nc(A))$ communicates plausibility, system
confidence, and certainty without requiring knowledge of the outcome.
Once the observed category $c_{\mathrm{obs}}$ is available,
Section~\ref{sec:native_scorecard} evaluates the same three components
in a post-event scorecard, expanding them to five metrics
(Table~\ref{tab:three_to_five}).

This framework bears structural resemblance to Dempster--Shafer
belief--plausibility
intervals \citep{Yager2005-at};
the pignistic transformation
\citep{Smets1990-nf} provides the decision-making conversion in
Dempster--Shafer theory just as
Section~\ref{sec:poss_prob_bridge} does here. The key difference is
that \ign{} has no direct Dempster--Shafer analogue: it quantifies
coverage gaps in the forecasting system
rather than evidential conflict.

We may reasonably ask: \textbf{why not just probabilities?} \textit{Is it worth the complexity?} The author submits this is so, though only if the glut of numbers is rendered in prose that is readable by all demographics. A probabilistic forecast assigns a single
number $p(A)$ to each outcome,
and these values must sum to unity.
A possibility distribution instead
provides the interval $[\nec(A),\, \poss(A)]$
that represents imprecision in the available
evidence. When evidence is thin (a
novel mesoscale pattern, a data-void region, a rapidly evolving
convective mode), this interval is wide, and the forecast admits it
cannot sharpen the estimate.
Probability forces a point value even when
the evidence does not support
one \citep{Dubois2007-dd}.

The ignorance signal \ign{} has no probabilistic counterpart. A
probability vector that sums to~1 cannot
express ``I don't know'', and any
admitted uncertainty must be allocated to
specific outcomes;
i.e., distributing confidence whether or not
such precision is warranted. A subnormal
possibility distribution reserves an explicit fraction of plausibility
as unassigned, signalling the system encounters conditions outside
its training or coverage.
This distinction allows downstream
decision-makers to
treat high-\ign{} forecasts differently
from low-\ign{} ones
(Section~\ref{sec:component_value}).

In severe-weather forecasting,
categorical language seen in outlook discussion
(e.g., ``\spc{MDT} risk is possible'')
already expresses plausibility rather than
calibrated frequency.
Possibility theory mathematically formalises
linguistics while preserving the ability to
derive probabilities when needed via the
conversion in Section~\ref{sec:poss_prob_bridge}.



\section{Converting Possibilities to Probabilities}%
\label{sec:pignistic_bridge}

Probabilistic verification scores familiar to
forecasters,
such as the logarithmic score (REF),
Brier score \citep{Brier1950-lc}, etc.,
cannot evaluate possibilities.
Values $\Pi \in \pi$ encode plausibility rather
than relative frequency,
and subnormal distributions are permissible
(i.e., a fully possible event is not enforced).
The conversion developed below transforms a
subnormal possibility distribution into an
$(n{+}1)$-category probability vector,
preserving the subnormality signal as explicit
ignorance rather than erasing through na\"ive
normalisation.
The term \textit{pignistic}
(from Latin \textit{pignus}, a wager) denotes
the probabilities one would assign if forced to
bet \citep{Hagedorn2009-pc};
the conversion shares this forced-betting
intuition.

\subsection{Possibility-to-Probability Conversion}%
\label{sec:poss_prob_bridge}

We seek specific characteristics when applying
probabilistic verification to possibilistic
forecasts.

\begin{workedexample}[Desiderata]
Any such conversion should satisfy:
\begin{enumerate}[leftmargin=2em, label=(\roman*)]
\item The result is a valid probability vector (sums to unity).
\item The ignorance signal $\ign$ is preserved, not erased.
\item For normal distributions ($\max\pi = 1$, $\ign = 0$), the
    conversion reduces to standard proportional normalisation.
\item Category ordering is preserved: if $\pi_i > \pi_j$ then
    $p_i > p_j$.
\end{enumerate}
\end{workedexample}

Peak possibility $m = \max(\pi)$ is \textit{commitment}: the
fraction of a unit plausibility budget that the system reserved for
its assessment. The remaining $1 - m$ is the
system's admission that it cannot fully explore outcome space.
The conversion sets this ignorance share aside
as an explicit ``unsure'' outcome \citep{Lawson2024-jc}.
The committed portion is then distributed
proportionally among the weather categories,
preserving their relative plausibility ranking.
A confident system ($m \approx 1$) stakes nearly
everything;
an uncertain one ($m \ll 1$) strands most mass
in the ignorance bin,
where it earns no verification credit but is a
neutral measure.

The pignistic
transformation \citep{Smets1990-nf,Sudano2015-rk} is valid for
Dempster--Shafer belief functions:
a similar approach to classical possibility
theory detailed in \citet{Yager2005-at}.
The full belief mass is redistributed across
focal elements to form a probability as
in \citet{Le_Carrer2021-by};
it does not reserve any mass for ignorance.
Hence all plausibility is converted,
regardless of how much evidential support it
carries.
For subnormal possibility distributions,
a system with $\ign = 0.5$ (admitting half its
budget is unsupported) would produce the same
probability vector as one with $\ign = 0$
(fully committed),
provided their raw shapes are proportional.
The conversion below avoids this by reserving
the ignorance mass as an explicit
$(n{+}1)$-th outcome,
ensuring subnormality directly affects
verification scores.
For normal distributions ($\ign = 0$),
both methods coincide.

\textbf{Formal procedure.} The conversion
proceeds in three steps:
\begin{enumerate}[leftmargin=2em]
\item \textbf{Reserve} ignorance mass:
    $p_{\mathrm{ign}} = \ign = 1 - \max(\pi)$.
\item \textbf{Distribute} remaining $(1 - \ign)$ proportionally:
    \[
    p_i = \frac{\pi_i \cdot (1 - \ign)}{\sum_j \pi_j}
    \]
\item \textbf{Append ignorance} $p_{\mathrm{ign}}$ as explicit
    outcome, yielding probability distribution over $n+1$
    categories that sums to unity.
\end{enumerate}

This satisfies all four desiderata: the result sums to one by
construction (i--ii); when $\ign = 0$, step~1 reserves nothing and the
formula reduces to $p_i = \pi_i / \sum_j\pi_j$ (iii); and
proportionality preserves ordering (iv).
The allocation sums possibility values rather than taking
maxima (violating max-additive Axiom~3); the author argues this is
acceptable as the goal is a probability vector.

\begin{figure*}[!t]
\centering
\includegraphics[width=0.85\textwidth]{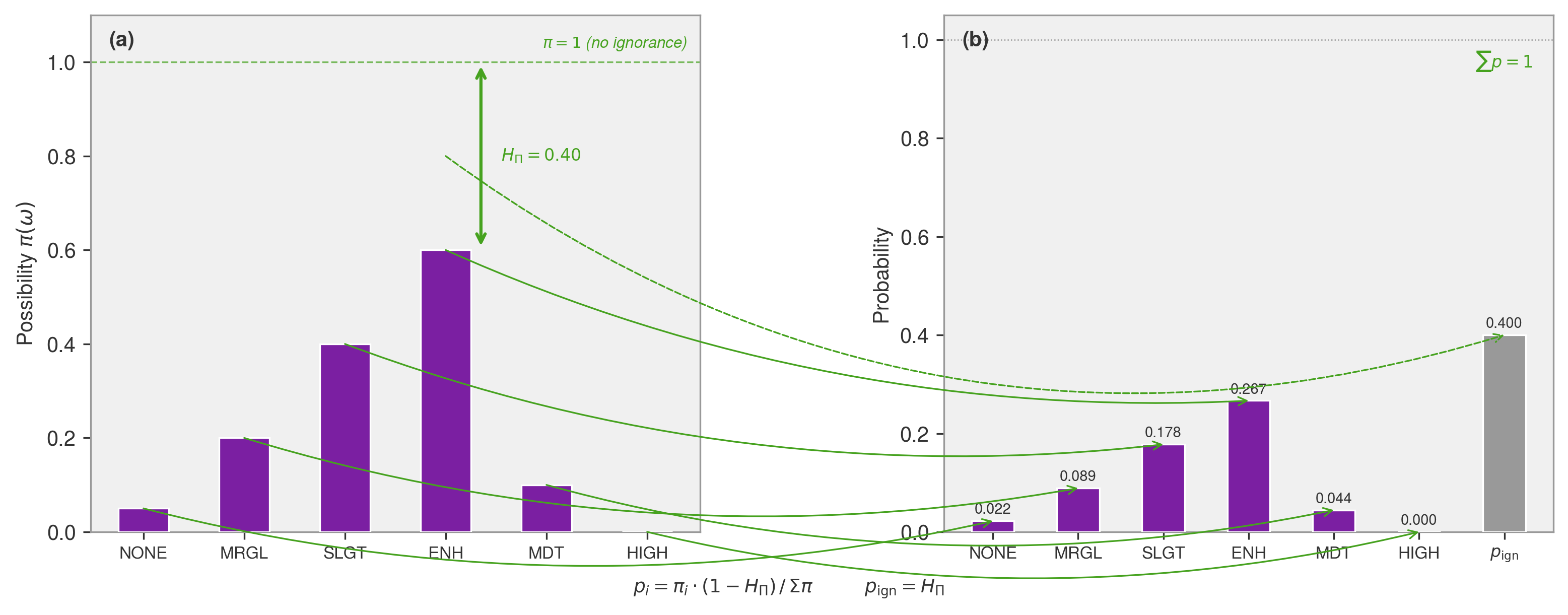}
\caption{Possibility-to-probability conversion
    applied to a subnormal distribution.
    \textbf{(a)}~Raw possibility values
    $\pi(\omega)$ for six SPC categories.
    \textbf{(b)}~Resulting $(n{+}1)$-category
    probability vector,
    with the grey bar representing the explicit
    ignorance outcome $p_{\mathrm{ign}} = \ign$.
    Arrows show proportional redistribution of
    the remaining $(1 - \ign)$ mass.
    The ignorance outcome absorbs probability
    mass that simple normalisation would have
    spread across all categories,
    preserving the subnormality signal.}
\label{fig:pignistic_bridge}
\end{figure*}

\begin{workedexample}[Worked example]
Suppose a possibilistic convective outlook produces
$\pi = (0.05,\; 0.2,\; 0.4,\; 0.6,\; 0.1,\; 0.0)$.
\begin{align*}
\textit{Input:}\quad
  & \ign = 1 - 0.6 = 0.4,\quad
    \textstyle\sum_j \pi_j = 1.35 \\
  & \quad (\text{40\% of plausibility budget unassigned}) \\[4pt]
\textit{Output:}\quad
  & \mathbf{p} = (0.022,\; 0.089,\; 0.178,\; 0.267, \\
  & \qquad 0.044,\; 0.0,\;
    \underbrace{0.400}_{\text{ign}}) \\[4pt]
\textit{Surprise:}\quad
  & \text{If \spc{ENH} obs: }
    S = {-}\log_2(0.267) = 1.91\;\text{bits} \\
  & S_{\mathrm{clim}} = {-}\log_2(0.06) = 4.06\;\text{bits} \\[4pt]
\textit{IG:}\quad
  & 4.06 - 1.91 = 2.15\;\text{bits gained}
\end{align*}

High \ign{} ($0.4$) inflated the ignorance outcome, reducing
$p(\spc{ENH})$ from what simple normalisation would give
($0.6/1.35 = 0.444$) to $0.267$. The forecast surprise ($1.91$~bits)
preserves the system's partial uncertainty.
Figure~\ref{fig:pignistic_bridge} visualises this transformation.
\end{workedexample}

\subsection{Information-Gain Decomposition}%
\label{sec:ig_decomposition}

Once possibilities have been converted to
probabilities, standard verification applies.
Define the \textbf{surprise} of a forecast at
the observed outcome as the log
score \citep{Shannon1948-nc}:
\begin{equation}
S(f,\, c_{\mathrm{obs}})
    = -\log_2 p_f(c_{\mathrm{obs}}) \quad \text{(bits)}
\label{eq:surprise}
\end{equation}
This appears throughout literature in
various forms
\citep{Good1952-st,Roulston2002-eq,Benedetti2010-sa,Weijs2010-hg, Peirolo2011-sl,Cover2012-di,Todter2012-ou}.
Low surprise means the forecast assigned high
probability to what actually happened;
the logarithm makes the score particularly
sensitive to confident misses on rare events.
To prevent $S \to \infty$ when
$p_f(c_{\mathrm{obs}}) = 0$,
converted probabilities are subjectively floored
at $\varepsilon = 0.01$ throughout;
Section~\ref{sec:bridge_walkthrough} provides
sensitivity analysis.

\textbf{Information gain} (IG) then measures the reduction in surprise
from replacing a baseline forecast $f_1$
(e.g., a previous model version,
or climatology) with forecast $f_2$:
\begin{equation}
\mathrm{IG}
    = S(f_1,\, c_{\mathrm{obs}})
    - S(f_2,\, c_{\mathrm{obs}})
\label{eq:ig}
\end{equation}
Positive IG indicates that the forecast reduced surprise relative to
the baseline \citep{Roulston2002-eq}. Equivalently, IG is the
difference in sample Kullback--Leibler divergences
from the observation to each
forecast \citet{Weijs2011-cf}.

Averaging Eq.~\eqref{eq:ig} over a verification
sample, the mean surprise decomposes into three
diagnostic terms \citep{Hersbach2000-yb}:
\begin{equation}
\overline{S}
    = \mathrm{UNC} - \mathrm{DSC} + \mathrm{REL}
\label{eq:ig_decomp}
\end{equation}
where \textbf{UNC} is the irreducible uncertainty (entropy of the base
rate), \textbf{DSC} is discrimination (can the system tell events
apart?), and \textbf{REL} is calibration error (are
stated probabilities honest?).
As in \citet{Lawson2024-bu},
the concept of ``sharpness'' is abbreviated by DSC
(discrimination) rather than RES (resolution) to avoid
confusion with horizontal grid resolution in
NWP.



When the baseline is climatology,
$\overline{S}(\mathrm{clim}) = \mathrm{UNC}$, and information gain
reduces to
$\mathrm{IG} = \mathrm{DSC} - \mathrm{REL}$: positive IG requires
discrimination to exceed calibration error. This decomposition is the
logarithmic counterpart of the Brier score's three-component breakdown
($\mathrm{BS} = \mathrm{REL} - \mathrm{RES} + \mathrm{UNC}$) but
operates in bits---additive, interpretable information units.

The analogue of DSC on the possibilistic scorecard
(Section~\ref{sec:native_scorecard}) is support margin~$\delta$:
whether an individual forecast-distribution
shape favoured the truth.
A system with consistently positive~$\delta$
will tend to exhibit high DSC after
conversion to probabilities, but the two are not interchangeable.
Together with critical success index
(CSI; \citealt{Jolliffe2003-xs})
of categories (Section XX),
the three quantities---$\delta$ (possibilistic,
per-forecast),
DSC (probabilistic, aggregate),
and CSI (categorical or binary)---triangulate
the forecast's ability to separate outcomes.

\begin{figure}[!b]
\centering
\includegraphics[width=\columnwidth]{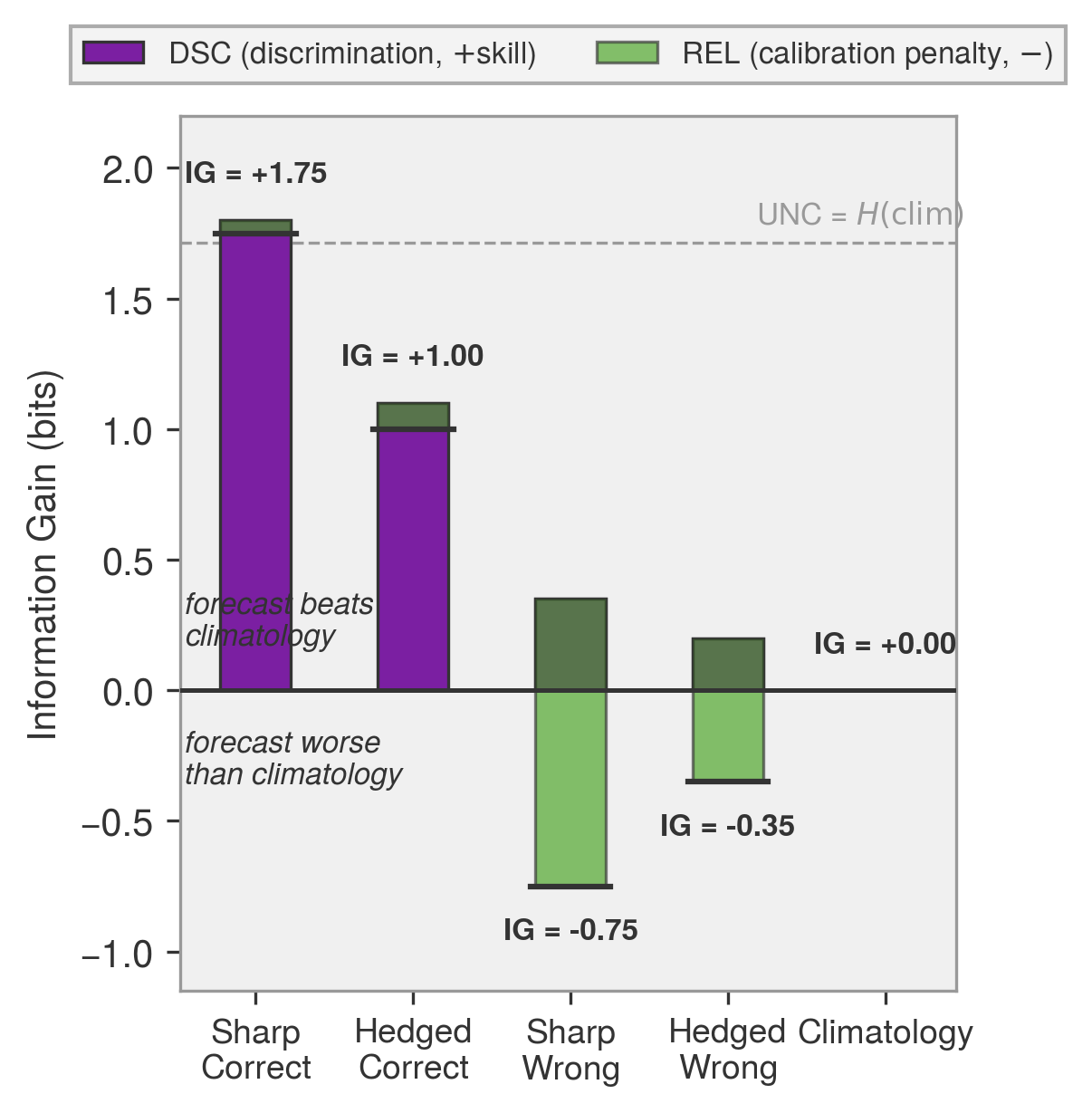}
\caption{Information-gain decomposition for five
    forecast archetypes (illustrative values).
    Purple bars show gross discrimination (DSC);
    semi-transparent green overlays show the
    reliability penalty (REL) consuming DSC
    from the top. Net
    $\mathrm{IG} = \mathrm{DSC} - \mathrm{REL}$ is marked by the
    horizontal tick on each bar; the zero baseline represents
    climatological performance. When REL exceeds DSC (Sharp Wrong,
    Hedged Wrong), the green extends below zero: the calibration tax
    consumed more skill than added through
    discrimination.
    DSC and REL are sample-aggregated quantities
    decomposed from a verification subset
    dominated by each forecast archetype.}
\label{fig:ig_decomp}
\end{figure}

Figure~\ref{fig:ig_decomp} illustrates the decomposition for five
forecast archetypes (distinct from the three worked scenarios of
Section~\ref{sec:worked_examples}).
The key contrast is between Sharp Correct
(nearly all discrimination retained, IG~$= +1.75$~bits) and Sharp
Wrong (the calibration penalty consumes the discrimination entirely
and extends below zero, IG~$= -0.75$~bits).
When comparing DSC and REL,
note Sharp-Wrong retains positive DSC despite
net information loss,
representing the salvaged benefit of ruling out
some outcomes,
which is then far outweighed by REL.
DSC measures whether the system produces different probability
assignments for different observed categories---the ability to
differentiate situations, regardless of direction. A sharp-wrong
system \textit{is} discriminating: it issues confident,
situation-specific forecasts that just happen to point the wrong way.
REL captures that directional error. Hedged Wrong shows less
discrimination and a smaller reliability penalty---a system that
hedges across wrong categories is penalised less harshly than one that
commits confidently to the wrong answer.

\subsection{Na\"ive normalisation erases information on ignorance}%
\label{sec:naive_normalisation}

Simple normalisation ($p_i = \pi_i / \sum_j \pi_j$) is the standard
route from possibilities to probabilities. It forces the result to sum
to one regardless of model confidence: a system with $\ign = 0.9$
(leaving 90\% of the unit-interval plausibility budget unassigned)
produces the same probability vector as one with $\ign = 0.1$,
provided their raw shapes are proportional.
The ignorance signal is erased.
Verification scores
applied to the normalised vector cannot distinguish a sharp, confident
forecast from a diffuse, hedged one---the answer to ``how would you
bet?'' is given unable to acknowledge
``I wouldn't bet at all'' \citep{Van-Schaeybroeck2016-sl}.

The conversion avoids this by deferring
probabilistic commitment for the ignorance portion. Mass equal to
$\ign$ is routed into an $(n{+}1)$-th outcome that can never be the
observed category. Uncertain forecasts are penalised because this
stranded mass earns no verification credit: it inflates the surprise
of the converted forecast in proportion to $\ign$. The more ignorant
the system, the more mass is stranded, and the larger the penalty.

\begin{workedexample}[Worked contrast]
Using $\pi = (0.05,\, 0.2,\, 0.4,\allowbreak\, 0.6,\, 0.1,\, 0.0)$
from Section~\ref{sec:poss_prob_bridge}:

\begin{center}
\small
\begin{tabular}{lcc}
\toprule
 & Simple norm.\ & Conversion \\
\midrule
$p(\spc{ENH})$ & $0.6/1.35 = 0.444$ & $0.267$ \\
$p_{\mathrm{ign}}$ & --- & $0.400$ \\
Surprise (bits) & $-\log_2(0.444) = 1.17$ & $-\log_2(0.267) = 1.91$ \\
\bottomrule
\end{tabular}
\end{center}

\noindent The $0.74$-bit gap is the verification
cost of ignoring admitted uncertainty.
Simple normalisation rewards the system for
confidence it never claimed.
\end{workedexample}

\textbf{Complementarity of \ign{} and IG.}
Possibilistic ignorance $\ign$ and the reduction
of ignorance measured by IG capture
complementary dimensions of forecast quality:
\begin{itemize}[leftmargin=2em]
\item \textbf{$\ign$ (possibilistic ignorance):} Measures the
    forecasting system's support strength---did
    the system know what was going on?
    Derived from the subnormality of the raw output
    (Eq.~\ref{eq:ignorance}).
\item \textbf{IG (information gain):}
    Measures forecast value---did the
    forecast reduce surprise about what happened? Derived from the
    probability-converted forecast evaluated against observations.
\end{itemize}

Both are needed because:
\begin{itemize}[leftmargin=2em]
\item A \textbf{sharp wrong} forecast has low \ign{} but negative IG.
\item A \textbf{hedged correct} forecast has
    high \ign{} but positive IG.
\item A \textbf{sharp correct} forecast has
    low \ign{} and high IG (ideal).
\item A \textbf{hedged wrong} forecast has high \ign{} and negative
    IG---the system knew it was unsure,
    and it still ruled the outcome implausible.
\end{itemize}

\FloatBarrier


\section{Possibilistic Verification}\label{sec:native_verification}

Converting possibilities to probabilities
(Section~\ref{sec:poss_prob_bridge}) is a lossy
transformation, compressing the possibility shape
into $n{+}1$ probabilities, with the entire subnormality
signal collapsed into a single ignorance bin.
Native possibilistic verification includes the raw
distribution, preserving the shape information otherwise
discarded. The metrics following now diagnose forecast
quality in possibility space.

All definitions use the notation of Section~\ref{sec:poss_primer}:
$\pi$ denotes the raw possibility distribution that may
or may not be subnormal (i.e., $\pi = \pi'$),
$m = \max_{\omega \in \Omega} \pi(\omega)$ is the
subnormality, and
$\ign = 1 - m$ is ignorance (Eq.~\ref{eq:ignorance}).

\subsection{Normalisation method}\label{sec:norm_protocol}

Regarding why some verification metrics herein
(Table~\ref{tab:raw_norm}) use the raw distribution
$\pi$ while others require the normalised form
$\pinorm = \pi / m$, the two
normalisation operations serve distinct purposes:

\begin{enumerate}[leftmargin=2em]
\item \textbf{Shape normalisation} ($\pi \to \pinorm$): Rescales so
    that $\max(\pi) = 1$, isolating the system's relative
    preferences among outcomes. Used for computing \Nc{}
    (Eq.~\ref{eq:cond_nec}) and the native scorecard
    (Section~\ref{sec:native_scorecard}).
\item \textbf{Possibility-to-probability conversion}
    ($\pi \to$ probability vector): Converts possibilities
    to probabilities with an explicit ignorance outcome
    (Section~\ref{sec:poss_prob_bridge}).
    Used for information-gain verification.
\end{enumerate}

\begin{table}[htbp]
\centering
\caption{Raw vs.\ normalised form used by each
    verification quantity.}
\label{tab:raw_norm}
\small
\renewcommand{\arraystretch}{1.15}
\begin{tabular}{@{}llcl@{}}
\toprule
Quantity & Symbol & Form & Reason \\
\midrule
\multicolumn{4}{@{}l}{\textit{Raw $(\pi)$}} \\[2pt]
\quad Ignorance              & $\ign$      & $\pi$    & Gap to normality \\
\quad Event possibility      & $\poss(A_T)$& $\pi$    & Upper bound \\

\multicolumn{4}{@{}l}{\textit{Normalised $(\pinorm)$}} \\[2pt]
\quad Depth-of-truth         & $\alpha^*$  & $\pinorm$ & Shape quality \\
\quad Diffuseness              & $\eta$      & $\pinorm$ & Shape spread \\
\quad Conditional necessity   & $\Nc(A)$    & $\pinorm$ & Certainty given coverage \\[4pt]
\bottomrule
\end{tabular}
\renewcommand{\arraystretch}{1.0}
\end{table}

\subsection{Five-Number Scorecard}\label{sec:native_scorecard}

Section~\ref{sec:three_component} introduced three
diagnostic components that characterise a forecast
\textit{before} the outcome is known:
possibility $\Pi$ (plausibility of an event),
ignorance $\ign$ (system confidence),
and conditional necessity $N_c$
(dominance among covered scenarios).
When ``truth" category $c_{\mathrm{obs}}$ is observed,
components can be evaluated against the truth.
Five summary statistics provide insight into these
three forecast metrics: how much support the truth
received ($\alpha^*$), how spread the forecast was
($\eta$), and whether the shape was net helpful
($\delta$). Ignorance $\ign$ carries forward
unchanged---it is a property of the forecast itself,
included to contextualise the other four metrics,
much as sharpness contextualises reliability in
probabilistic verification.
Table~\ref{tab:three_to_five} maps the three pre-event components to
the five post-event metrics.

\begin{workedexample}[Scorecard desiderata]
A possibilistic scorecard should:
\begin{enumerate}[leftmargin=2em, label=(\roman*)]
\item Cover three uncertainty aspects of the forecast:
    possibility, ignorance, and conditional necessity.
\item Separate shape from confidence: factor out the commitment
    $m$ so shape quality and confidence are evaluated
    independently.
\item Separate support from spread: distinguish ``did
    the system favour the truth?'' from ``was the
    forecast sharp?'' (sharp-wrong and hedged-correct
    forecasts fail in different ways).
\item Include a dominance indicator: when assigning
    possibility to observed event $\hat{c}$,
    was $\hat{c}$ the system's unique top pick or
    merely one of several equally plausible categories?
\end{enumerate}
\end{workedexample}

\begin{table}[htbp]
\centering
\caption{From three components to five scorecard metrics. Left: the
    three diagnostic components characterise a forecast before the
    outcome is known (Section~\ref{sec:three_component}). Right: once
    $c_{\mathrm{obs}}$ is available, five metrics evaluate the same
    components. The possibility component expands to three metrics;
    ignorance carries forward as forecast context.}
\label{tab:three_to_five}
\small
\begin{tabular}{lll}
\toprule
Component & Pre-event & Post-event scorecard \\
\midrule
Possibility & $\poss(A)$, $\pinorm(\omega)$
    & $\alpha^*$, $\eta$, $\delta$ \\
Ignorance & $\ign$   & $\ign$ \\
Necessity & $\Nc(A)$  & $\Nc^*$ \\
\bottomrule
\end{tabular}
\end{table}

Five summary statistics evaluate each
forecast--observation pair, where $c_{\mathrm{obs}}$
denotes the issued SPC category (standing in for
"verification); for the point of interest,
$K = |\Omega|$ is the number of categories
(here, $K = 6$).
Table~\ref{tab:scorecard_defs} lists the definitions;
the equations follow.

\begin{table}[htbp]
\centering
\caption{Five-number scorecard definitions. The Component column
    groups metrics by the three diagnostic components of
    Table~\ref{tab:three_to_five}.}
\label{tab:scorecard_defs}
\small
\begin{tabular}{clcl}
\toprule
Symbol & Name & Component & Eq.\ \\
\midrule
$\alpha^*$ & Depth-of-truth & Possibility
    & \eqref{eq:alpha_star} \\
$\eta$ & Diffuseness & Possibility
    & \eqref{eq:ns} \\
$\delta$ & Support margin & Possibility
    & \eqref{eq:rg} \\
$\ign$ & Ignorance & Ignorance
    & \eqref{eq:ignorance} \\
$\Nc^*$ & Cond.\ necessity of truth & Necessity
    & \eqref{eq:nc_star} \\
\bottomrule
\end{tabular}
\end{table}

\textbf{Depth-of-truth} ($\alpha^*$):
\begin{equation}
\alpha^* = \pinorm(c_{\mathrm{obs}}) \label{eq:alpha_star}
\end{equation}
$\alpha^* = 1$ when the truth coincides with the
system's peak category, $\alpha^* = 0$ when the
system assigns zero possibility to the observed
outcome, and intermediate values arise whenever
the truth is
plausible but not the system's top pick.

\textbf{Diffuseness} ($\eta$) measures the spread of
the normalised distribution shape,
a parallel of non-specificity in \citet{Klir1995-va},
computed here as an arithmetic mean that uses a
baseline of uniform sampling across categories
(echoing the na\"ive 50:50 in the leading paragraph):

\begin{equation}
\eta = \frac{1}{K} \sum_{c=1}^{K} \pinorm(c) \label{eq:ns}
\end{equation}
ranging from $1/K$ (sharp) to $1$ (uniform).

\textbf{Support margin} ($\delta$):
\begin{equation}
\delta = \alpha^* - \eta \label{eq:rg}
\end{equation}
Positive $\delta$ indicates the system assigned more
possibility to the truth than to an average category;
negative $\delta$ \textit{vice versa}.
The term \textit{support margin} reflects this
interpretation: $\delta$ measures the margin of excess
support (normalised possibility) that the truth
received relative to the categorical average.
It is the per-forecast possibilistic counterpart of
DSC (Section~\ref{sec:ig_decomposition}),
which measures separation of events from non-events.
The name avoids clashing with either ``resolution''
(already replaced by DSC to prevent confusion with
NWP grid resolution) or ``discrimination''
(reserved for DSC).

Further conceptual connection with Brier-type
resolution lends interpretation of ``how well
forecasts separate events from climatology",
or detection of signal over noise.
When evaluating deterministic/categorical forecasts,
analogy lies in the Peirce Skill Score
(hit rate minus false-alarm rate;
\citealt{Wilks2011-vw}) that also measures
discrimination above baseline.

\textbf{Ignorance} ($\ign$), carried forward from the pre-event
framework:
\begin{equation}
\ign = 1 - m = 1 - \max_{\omega \in \Omega} \pi(\omega)
\end{equation}
which defined the concept of \textit{commitment} $m$
as the complement of Eq.~\ref{eq:ignorance}.
This is the only scorecard metric
that uses the raw distribution; the others use the normalised
shape $\pinorm$.

\textbf{Conditional necessity of truth} ($\Nc^*$):
\begin{equation}
\Nc^* = \Nc(c_{\mathrm{obs}})
    = 1 - \max_{\omega \neq c_{\mathrm{obs}}} \pinorm(\omega)
\label{eq:nc_star}
\end{equation}
$\Nc^*$ serves as a \textit{dominance margin}:
it measures how strongly the truth dominated
alternatives in the normalised shape,
analogous to ensemble spread (i.e., diversity of
Monte Carlo simulations).
When observed category $c_obs$ is not the unique
peak of the normalised distribution, then
$\Nc^* = 0$, since $\pi = \pinorm$);
positive values quantify the margin by which truth
led the next-highest category.

Hence $\Nc^*$ provides diagnostic information only
on the subset of forecasts where $\alpha^* = 1$
(i.e., where truth coincides with the peak category),
distinguishing dominant peaks from contested ones.
When $\alpha^* < 1$, $\Nc^*$ is structurally zero.
This is an application of
Eq.~\eqref{eq:cond_nec} evaluated at the observation.

\medskip
\noindent\textbf{\textsf{\color{rationalecolor}Scorecard descriptions}}
\smallskip

\noindent
\renewcommand{\arraystretch}{1.25}
\begin{tabularx}{\columnwidth}{%
    >{\columncolor{rationalecolor!10}}c X}
$\boldsymbol{\alpha^*}$ & \textbf{Depth-of-truth.} How much normalised
    possibility did the observed category receive? High $\alpha^*$ indicates the system recognised outcome $c_obs$. \\
$\boldsymbol{\eta}$ & \textbf{Diffuseness.} How spread out was the
    normalised distribution? Analogue of ensemble spread or sharpness. \\
$\boldsymbol{\delta}$ & \textbf{Support margin.} Did the system give
    more support to the truth than to an average category?
    Analogue of DSC (REL) Brier-type resolution. \\
$\boldsymbol{\ign}$ & \textbf{Ignorance.} How much of the plausibility
    budget did the system leave unassigned? $\ign = 0$: fully
    confident; $\ign \to 1$: near-total ignorance. \\
$\boldsymbol{\Nc^*}$ & \textbf{Cond.\ necessity of truth.} Was truth a clear winner? A conceptual merging of recognition and conviction. \\
\end{tabularx}
\renewcommand{\arraystretch}{1.0}

\begin{rhoenv}[frametitle={Definition (Five-Number Possibilistic Scorecard)}]
Given a subnormal possibility distribution $\pi$ over $\Omega$ with
$m = \max\pi > 0$, normalised form $\pinorm = \pi/m$, and observed
outcome $c_{\mathrm{obs}}$:
\begin{align*}
\alpha^* &= \pinorm(c_{\mathrm{obs}}) &&\in [0,\,1] \\
\eta &= \tfrac{1}{K}\textstyle\sum_{c}\pinorm(c) &&\in [1/K,\,1] \\
\delta &= \alpha^* - \eta &&\in [-(1{-}1/K),\,1{-}1/K] \\
\ign &= 1 - m &&\in [0,\,1) \\
\Nc^* &= 1 - \max_{\omega \neq c_{\mathrm{obs}}} \pinorm(\omega) &&\in [0,\,1]
\end{align*}
\end{rhoenv}

The scorecard is constructive rather than axiomatic:
the five metrics are chosen to separate four
diagnostic dimensions---support (did the shape
favour the truth?), spread (how diffuse was it?),
commitment (how much of the plausibility budget was
spent?), and dominance (did the truth lead
its rivals?)---that collapse under any single-number summary.
The four scorecard desiderata are satisfied: all three components are
covered~(i); four metrics use the normalised
shape~$\pinorm$ while only
$\ign$ uses raw~$\pi$, separating shape from commitment~(ii);
$\alpha^*$ and~$\delta$ capture support while~$\eta$ captures
spread~(iii); and~$\Nc^*$ provides the dominance indicator~(iv).

Each metric is well-defined for a single forecast--observation pair;
overbar notation ($\overline{\alpha^*}$, $\overline{\eta}$, etc.)
denotes aggregation over the verification sample.
All five metrics treat $\Omega$ as an unordered set---they evaluate
whether the forecast pointed at the right category, not whether it was
close in an ordinal sense. For ordinal domains with well-defined
distances, a distance-sensitive score could supplement the scorecard.
The multi-metric design resists trivial gaming strategies: for
instance, flooring all values at $\varepsilon$ to avoid $\alpha^* = 0$
inflates $\eta$ on every forecast, yielding a net-negative effect
across the verification sample
(Section~\ref{sec:scorecard_comparison}).

Amongst risk-based verification rules,
\textit{propriety} is a gold-standard property that
resists hedging: honest reporting should maximise
the forecaster's expected score.
(Alternatively: the forecasts should not be smoothed
to ``play it safe").
The logarithmic score in the probabilistic facet
(Eq.~\ref{eq:surprise}) satisfies this criterion
(REFS). For the native scorecard,
propriety remains an open question: the five metrics
are descriptive diagnostics, not a single loss
function to buttress optimisation. The multi-metric
coupling nevertheless offers insight and discourages
gaming. Indeed, preliminary investigation showed that
improving one metric while holding the others fixed
is generally not possible (e.g., broadening
support to raise~$\alpha^*$ inflates~$\eta$ and
penalises~$\delta$). Establishing formal propriety for
possibilistic scorecards is left for future work.

\subsection{Categorical Verification Scores}%
\label{sec:cat_scores}

\begin{figure}[t]
\centering
\includegraphics[width=\columnwidth]{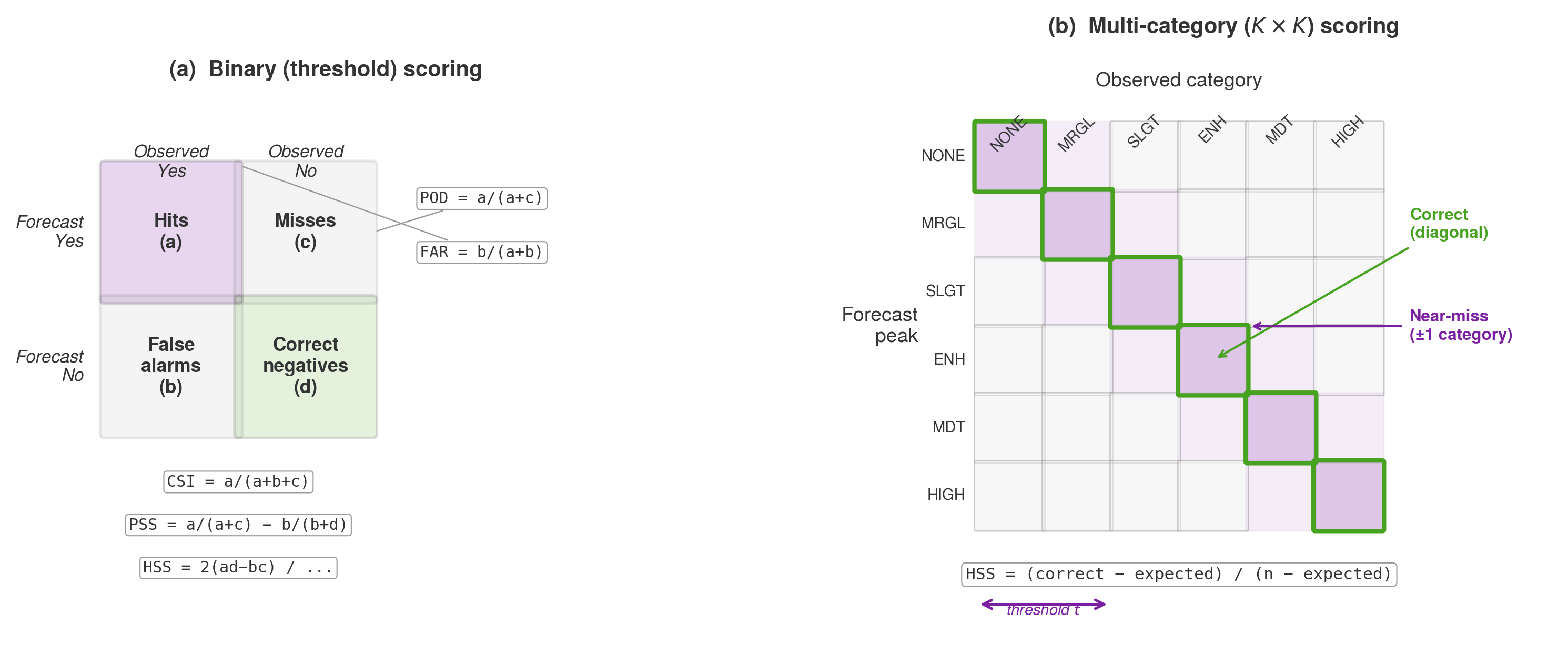}
\caption{Contingency-table schematic.
    \textbf{(a)}~Binary $2 \times 2$ table at a single
    severity threshold, annotated with the five categorical
    scores (Eqs.~\ref{eq:pod}--\ref{eq:hss}).
    \textbf{(b)}~Extension to the $K \times K$
    multi-category setting (SPC categories). Green boxes
    mark the diagonal (correct peak category);
    shaded cells mark $\pm 1$ near-misses. HSS
    (Eq.~\ref{eq:hss_kxk}) summarises full-table agreement
    beyond chance.
    The dashed bracket illustrates how each severity
    threshold~$t$ collapses the $K \times K$ table into a
    binary problem.}
\label{fig:contingency_schematic}
\end{figure}

The possibilistic scorecard evaluates distributional shape;
a complementary \textbf{categorical facet} reduces each
distribution to its peak category
$\hat{c} = \arg\max \pi(\omega)$ (ties broken by severity)
and applies threshold-based contingency scores familiar to
operational meteorologists
\citep{Jolliffe2003-xs,Wilks2011-vw}.
For each severity threshold~$t$ (\spc{MRGL}+, \spc{SLGT}+,
\ldots, \spc{HIGH}), the forecast issues ``yes'' if
$\hat{c} \geq t$ and ``no'' otherwise; the observation is
``yes'' if $c_{\mathrm{obs}} \geq t$.
This binary classification yields a $2 \times 2$ contingency
table with hits~($a$), false alarms~($b$), misses~($c$),
and correct negatives~($d$)
(Figure~\ref{fig:contingency_schematic}a).
Five scores evaluate each threshold:
\begin{align}
\mathrm{POD}  &= \frac{a}{a + c}
    &&\text{(detection)} \label{eq:pod} \\
\mathrm{FAR}  &= \frac{b}{a + b}
    &&\text{(false-alarm ratio)} \label{eq:far} \\
\mathrm{CSI}  &= \frac{a}{a + b + c}
    &&\text{(critical success)} \label{eq:csi} \\
\mathrm{PSS}  &= \frac{a}{a+c} - \frac{b}{b+d}
    &&\text{(Peirce skill)} \label{eq:pss} \\
\mathrm{HSS}  &= \frac{2(ad - bc)}{(a{+}c)(c{+}d) + (a{+}b)(b{+}d)}
    &&\text{(Heidke skill)} \label{eq:hss}
\end{align}

POD alone rewards over-forecasting; FAR alone rewards
under-forecasting; CSI balances both but is base-rate
dependent.
PSS (also known as the Hanssen--Kuipers discriminant or
true skill statistic; \citealt{Wilks2011-vw}) equals
$\mathrm{POD} - \mathrm{POFD}$,
where $\mathrm{POFD} = b/(b+d)$ is the probability of
false detection.
PSS is \textit{equitable}: random and constant forecasts
score zero, and it is independent of event frequency,
making it well-suited for rare-event verification.
HSS extends naturally to the full $K \times K$ confusion
matrix (Figure~\ref{fig:contingency_schematic}b):
\begin{equation}
\mathrm{HSS}_{K\times K}
  = \frac{n_{\mathrm{correct}} - n_{\mathrm{expected}}}%
         {n - n_{\mathrm{expected}}}
\label{eq:hss_kxk}
\end{equation}
where $n$ is the sample size and $n_{\mathrm{expected}}$ is
the number of correct forecasts expected by chance under
marginal-frequency matching.

\paragraph{Why these five scores?}
The selection balances three criteria.
First, POD, FAR, and CSI are the standard threshold
contingency triad that underpins the \citet{Roebber2009-rv}
performance diagram; retaining them allows direct comparison
with deterministic verification practice.
Second, PSS is the categorical analogue of the possibilistic
support margin~$\delta$
(Section~\ref{sec:native_scorecard}): both quantify
discrimination above a na\"ive baseline (climatology for
$\delta$; random forecasting for PSS).
Third, HSS aggregates over the full $K \times K$ table,
providing a single chance-corrected summary of multi-category
agreement analogous to mean IG in the probabilistic facet.
Together the five scores span detection, false-alarm control,
balanced accuracy, equitable discrimination, and
multi-category skill.

\FloatBarrier

\subsection{Diagnostic Diagrams}\label{sec:diag_diagrams}

The diagnostic diagrams in this section and the
aggregate figures in
Sections~\ref{sec:tripartite_value}--\ref{sec:worked_examples}
share a single synthetic reforecast
($n = 800$ active days) that mimics SPC categorical
forecasts in a possibilistic point-location manner.
The sample size represents approximately eight years
of ${\sim}100$ active Great Plains Day-1 convective
outlooks per season, assuming a stationary
verification climate.

To create the synthetic data for each ``day",
an observed category is drawn from approximate SPC
climatological frequencies
(\spc{NONE}~60\%, \spc{MRGL}~18\%,
\spc{SLGT}~12\%, \spc{ENH}~6\%,
\spc{MDT}~3.2\%, \spc{HIGH}~0.8\%).
A forecast peak category is selected:
correct with probability $p_{\mathrm{correct}}(c)$
(ranging from 82\% for \spc{NONE} to 18\% for
\spc{HIGH}), otherwise shifted to a near-miss
category ($\pm 1$ categories with probabilities
$0.4$; $\pm 2$ with $0.1$).
The possibility distribution is generated as an
exponential decay from the
peak:

\begin{equation}
\pi(c) = \exp\!\bigl(-\sigma\,|c - c_{\mathrm{peak}}|\bigr) + U(0,\, 0.03)
\label{eq:synth_gen}
\end{equation}

where $\sigma$ is a category-dependent spread
parameter (mean $\bar\sigma
\approx 2.6$ for \spc{NONE},
$\approx 0.7$ for \spc{HIGH}) and the
uniform noise term breaks symmetry. Finally,
the distribution is rescaled so that its peak value
equals $1 - h$, where $h$ is a randomly drawn
ignorance level that increases with category rarity:
mean ignorance is approximately $0.06$ for
\spc{NONE} (very confident) and $0.52$ for \spc{HIGH} (substantially
uncertain), reflecting the operational reality that
rare-event forecasts
carry more uncertainty.

These choices produce a dataset with category-dependent accuracy and
confidence gradients mimicking a realistic but imperfect possibilistic
system. At $n = 800$, the sample yields approximately 6 \spc{HIGH}
days and 26 \spc{MDT} days, reflecting the inherent sample-size
challenge in rare-event verification.

\begin{figure}[t]
\centering
\includegraphics[width=\columnwidth]{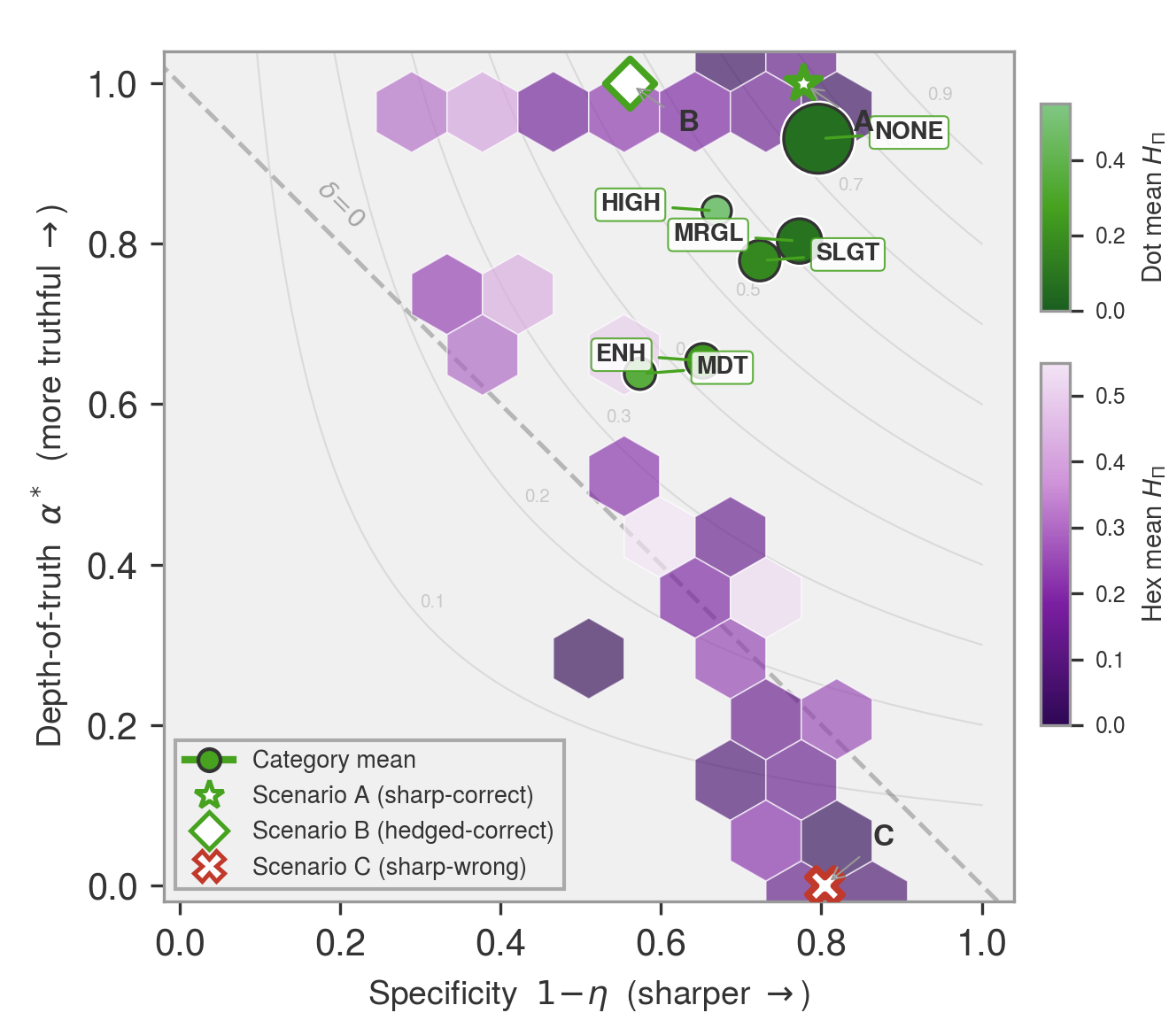}
\caption{Possibilistic performance diagram for the
    800-day synthetic reforecast,
    inspired by \citet{Roebber2009-rv}.
    Hexagons bin in Cartesian space where colour
    denotes mean $\ign$ (dark purple = confident,
    pale = uncertain).
    Contours join a joint skill
    $S = \alpha^* \times (1 - \eta)$,
    the possibilistic analogue of CSI,
    where $S$ is maximised in the upper right
    (sharp and truthful).
    The dashed diagonal traces $\delta = 0$,
    a break-even threshold;
    above this line the forecast discriminated and
    offered more support than a categorical average.
    The green circles show category means
    (dot size $\propto$ sample count),
    showing forecast quality as a function of
    observed category, tracing the progression from
    easy categories (upper right) to rare ones
    (lower left).
    Stars mark worked scenarios from
    Section~\ref{sec:worked_examples}
    (green edge = hit, red = miss).
    Lower right is the worst failure mode
    (sharp and wrong).}
\label{fig:performance_diagram}
\end{figure}

Figure~\ref{fig:performance_diagram} provides a
diagnostic view of the five scorecard metrics.
Hexagonal bins aggregate forecast--observation
pairs by specificity $1 - \eta$ ($x$-axis) and
depth-of-truth $\alpha^*$ ($y$-axis);
bin colour encodes mean ignorance $\ign$.
The analogy with the \citet{Roebber2009-rv}
performance diagram is deliberate:
specificity plays the role of success ratio,
$\alpha^*$ plays the role of POD, and the
$S = \alpha^* \times (1 - \eta)$ hyperbolas play
the role of CSI contours---all three axes are
maximised in the upper right. The key addition is
ignorance, encoded as bin colour,
which has no categorical analogue.

Sample are evident in a diagonal band from upper
left to lower right, dominated by \spc{NONE} days
(dark purple: confident). Green dots
mark mean values stratified by their (labelled)
category. Both axes of
Figure~\ref{fig:performance_diagram} operate on the
normalised shape $\pinorm$: two forecasts with
identical shapes but different commitment levels
land at the same position. Commitment is visible
only through bin colour and less precise than
Cartesian location.

\begin{figure}[t]
\centering
\includegraphics[width=\columnwidth]{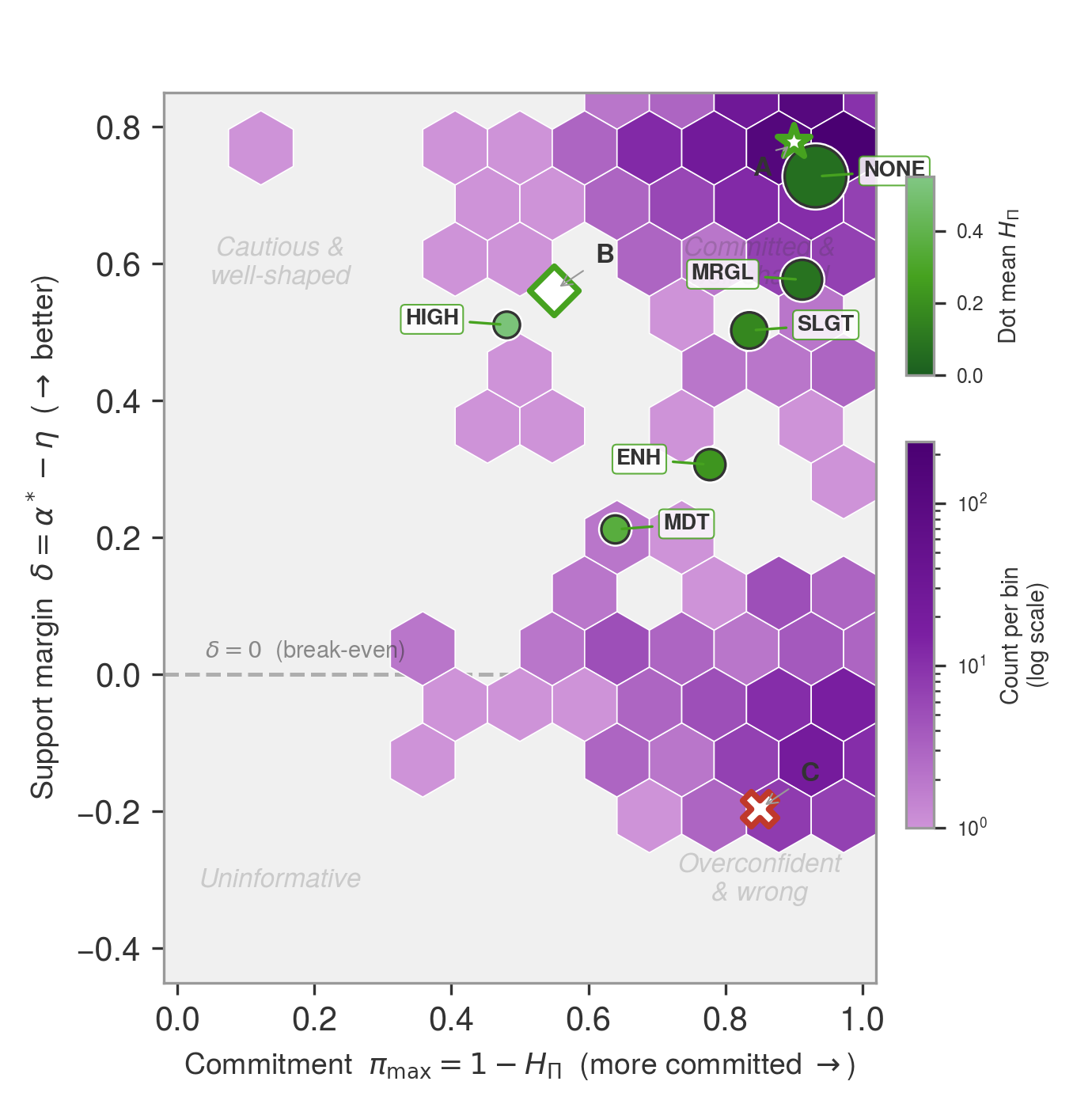}
\caption{Commitment--support-margin diagram for the
    same synthetic dataset.
    The x-axis shows commitment
    $m = \pi_{\max} = 1 - \ign$
    (how much the forecaster staked).
    The y-axis gives margin of support
    $\delta = \alpha^* - \eta$
    (did the shape favour truth?).
    These axes are orthogonal:
    $\pi_{\max}$ is the raw scale of $\pi$;
    $\delta$ is the quality of the normalised
    shape $\pi'$.
    Hex bins are coloured by count (density).
    Green dots mark category means as in
    Fig.~\ref{fig:performance_diagram}.
    Starred labels mark
    Section~\ref{sec:worked_examples} scenarios
    as in Fig.~\ref{fig:performance_diagram}.}
\label{fig:commitment_diagram}
\end{figure}

Consequently, Fig.~\ref{fig:commitment_diagram}
promotes commitment to a primary axis, plotting
$m = 1 - \ign$ against $\delta$. These axes
are orthogonal: $\pi_{\max}$ is the raw
\emph{scale} of $\pi$; $\delta$
is the \emph{shape quality} of $\pinorm$. Two clusters of high-count
hexagons emerge: one in the upper right
(confident and correct---mostly
\spc{NONE} days) and one in the lower right (confident and
wrong---forecasts that committed heavily but missed). The lighter band
between them contains the hedged severe-weather forecasts. The green
dots (category means) all stay above $\delta = 0$ for all
categories on average---even rare-event forecasts discriminate in
aggregate, though with lower commitment.
Scenario~C (sharp-wrong) sits towards the
overconfident quadrant;
Scenario~B (hedged-correct) sits towards the
cautious/well-shaped region---a distinction invisible in the
shape--quality diagram alone.



\section{Three-Component Value}\label{sec:tripartite_value}

Each component of the five-number scorecard
(Table~\ref{tab:three_to_five}) must provide
diagnostic information the others cannot. This section first examines
the independent value of the three components---possibility (shape
quality), ignorance (system confidence), and conditional necessity
(dominance)---and returns to connect possibilistic
forecasts to probabilistic and categorical verification.

\subsection{Value of each component}\label{sec:component_value}

\paragraph{Possibility: does the shape point at the truth?}

The possibility elements ($\alpha^*$, $\eta$, $\delta$) capture
whether the system's distributional shape favours the observed outcome.
Its value reduces to asking if the system's
support margin $\delta$ exceeds that of a na\"ive baseline.
Accordingly, comparing $\delta$ against climatology
or persistence is a possibilistic
analogue of Brier skill-score comparison
(Section~\ref{sec:limitations}. For SPC outlooks, climatology
peaks at the most frequent category and assigns near-zero possibility to
rare outcomes like \spc{HIGH}; a skilful possibilistic system should
outperform this baseline on days when \spc{ENH} or higher categories
verify, exhibiting positive $\delta$ across the verification sample.

\paragraph{Ignorance: self-reported uncertainty}

The ignorance component ($\ign$) carries forward from the pre-event
framework as a neutral forecast property
rather than a verification metric
(Table~\ref{tab:three_to_five}).
The admittance of ignorance is useful \textit{per se};
there is further benefit when coupled with
other scorecard metrics.
For example, if high-\ign{} forecasts tend to have lower
mean $\delta$ than low-\ign{} forecasts, then $\ign$ tracks actual
forecast difficulty---the possibilistic analogue of the
\textit{spread--skill relationship} in ensemble forecasting, where high
ensemble spread should correspond to higher forecast error.

There is overlap here with
``predicting the predictability'' \citep{Coleman2024-hf};
implications are critical for insight
into system design.
If \ign{} does \emph{not} track difficulty,
the subnormality signal may be uninformative,
hence normalisation is costless.
There is little reason to enforce
$\max(\pi) = 1$ to simplify the framework. If \ign{} \emph{does}
track difficulty, normalisation destroys operationally valuable
information: power to distinguish confident forecasts from hedged ones.

\paragraph{Conditional necessity: does claimed certainty verify?}

When the system assigns high conditional necessity to an outcome, that
outcome should verify more often than the base rate. This is the
possibilistic analogue of \textit{reliability conditional on
confidence}: ensemble forecast calibration answers
the prompt, ``when the ensemble
gives ${>}80\%$ chance of severe weather,
does it verify ${>}80\%$ of the time?''
Here the question becomes:
when $\Nc(\hat{c}) > \tau$ for
threshold $\tau$, does the conditional hit rate exceed climatology?
Note the shift: here we condition on the forecast's own peak category
$\hat{c}$ rather than the observation, since reliability is a
pre-verification property of the issued forecast.
In sum, high-necessity forecasts identify genuinely confident
predictions---not just lucky ones.

\begin{figure}[t]
\centering
\includegraphics[width=0.75\columnwidth]{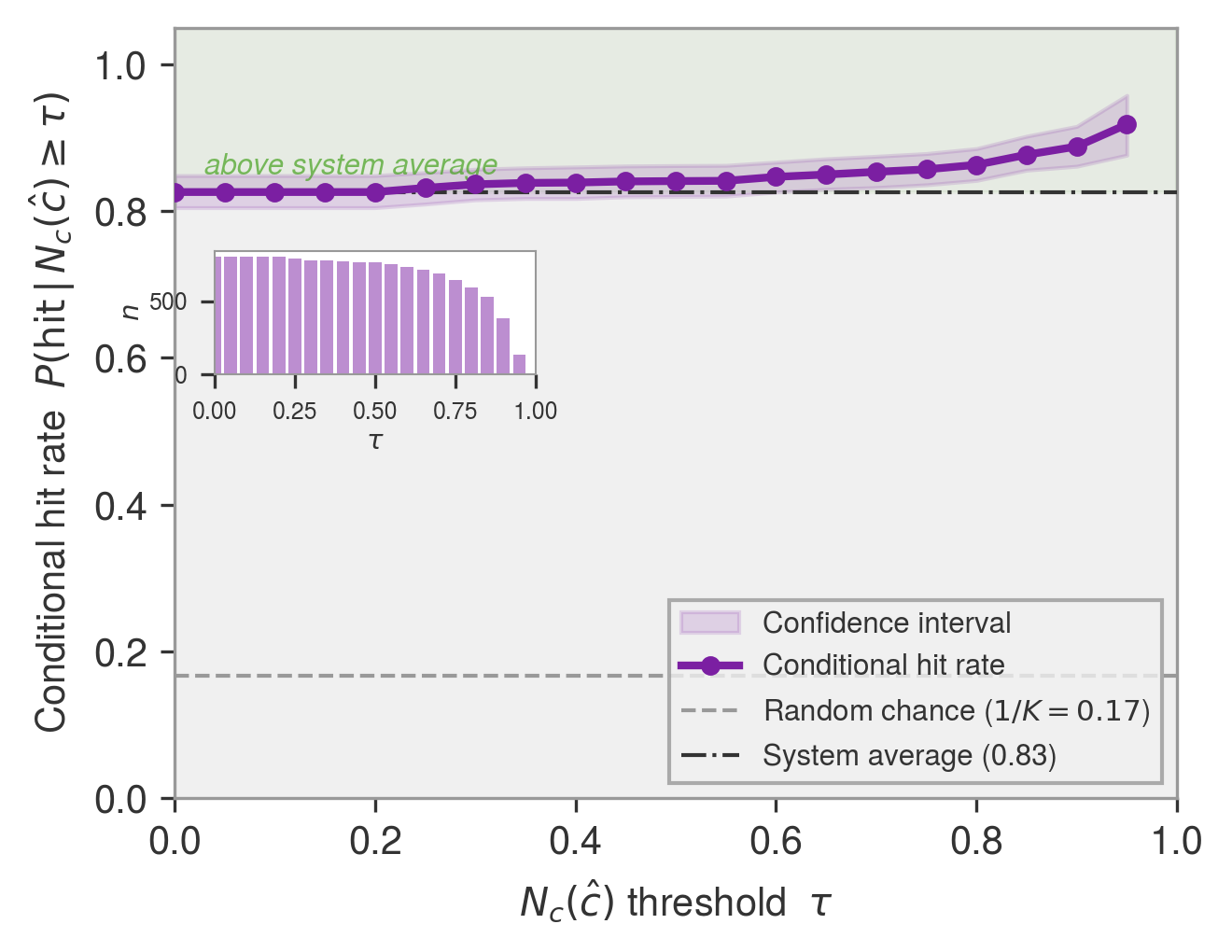}
\caption{Conditional hit rate vs.\ $\Nc(\hat{c})$ threshold for the
    synthetic reforecast. At each $\tau$, only forecasts with
    $\Nc(\hat{c}) \geq \tau$ are retained; the purple curve shows
    their hit rate. Baselines: random chance ($1/K$, dashed) and
    unconditional system accuracy (dot-dashed). Shaded band: notional
    confidence interval. Inset: declining sample size at high $\tau$.
    The near-flat curve reflects the synthetic system's high baseline
    accuracy; the uptick at $\tau > 0.8$ confirms
    that $\Nc(\hat{c})$ adds confidence signal.}
\label{fig:reliability}
\end{figure}

Figure~\ref{fig:reliability} tests this by plotting the conditional hit
rate against the $\Nc(\hat{c})$ threshold~$\tau$: at each $\tau$, only
forecasts whose peak category has $\Nc(\hat{c}) \geq \tau$ are included.
The curve is nearly flat because the synthetic system already has a high
baseline accuracy (${\sim}83\%$). The modest tick-up at high
thresholds ($\tau > 0.8$) confirms the most
confident forecasts (where the peak category dominates 
alternatives) verify more reliably, so $\Nc(\hat{c})$ carries a genuine
confidence signal. With real data or a weaker baseline system, this
separation would be more pronounced.

\subsection{Three Verification Facets}\label{sec:three_lanes}

The paper proposes three expressions for forecast uncertainty
and five metrics to evaluate forecast quality.
The \textbf{categorical facet} reduces each distribution to
its peak category and applies threshold-based contingency
scores---POD, FAR, CSI, PSS, and HSS---defined in
Section~\ref{sec:cat_scores}
(Eqs.~\ref{eq:pod}--\ref{eq:hss_kxk}).
Because verification is threshold-based, a peak forecast of
\spc{MDT} when \spc{HIGH} verifies registers as a hit at
all thresholds up to \spc{MDT}+ and a miss only at
\spc{HIGH}---partial credit for near-misses in severity.
The categorical product asks, ``did the peak-category
forecast identify the correct severity category?'', then
discards all distributional information: a correct peak
says nothing about sharpness, hedging, or confidence.

The \textbf{probabilistic facet}
(IG via probability conversion) asks
``was the probabilistic forecast informative
(reduce surprise)?''
The conversion is lossy---it compresses
the full possibility shape into $n{+}1$
probabilities, with subnormality collapsed into a single ignorance bin.

The native \textbf{possibilistic} paradigm asks
``was the possibility distribution well-shaped, sharp, and honest?''
It has fleeting connection to probabilistic skill:
a well-shaped distribution might still give
poor probabilities after conversion.

Together, the three sets of metrics can expose
failures that any single facet may miss.
A forecast with correct peak category but flat
possibility and poor IG is a broken clock
that happens to be right.
A forecast with strong IG but high ignorance
and low $\Nc^*$ was lucky after conversion
due to hedging in the raw system distribution.
A forecast with sharp possibility
and high $\Nc^*$ but poor IG may indicate
distortion introduced by the conversion---the
probability derivation is spuriously transforming
the signal through translation,
compression, etc.

\begin{figure}[t]
\centering
\includegraphics[width=\columnwidth]{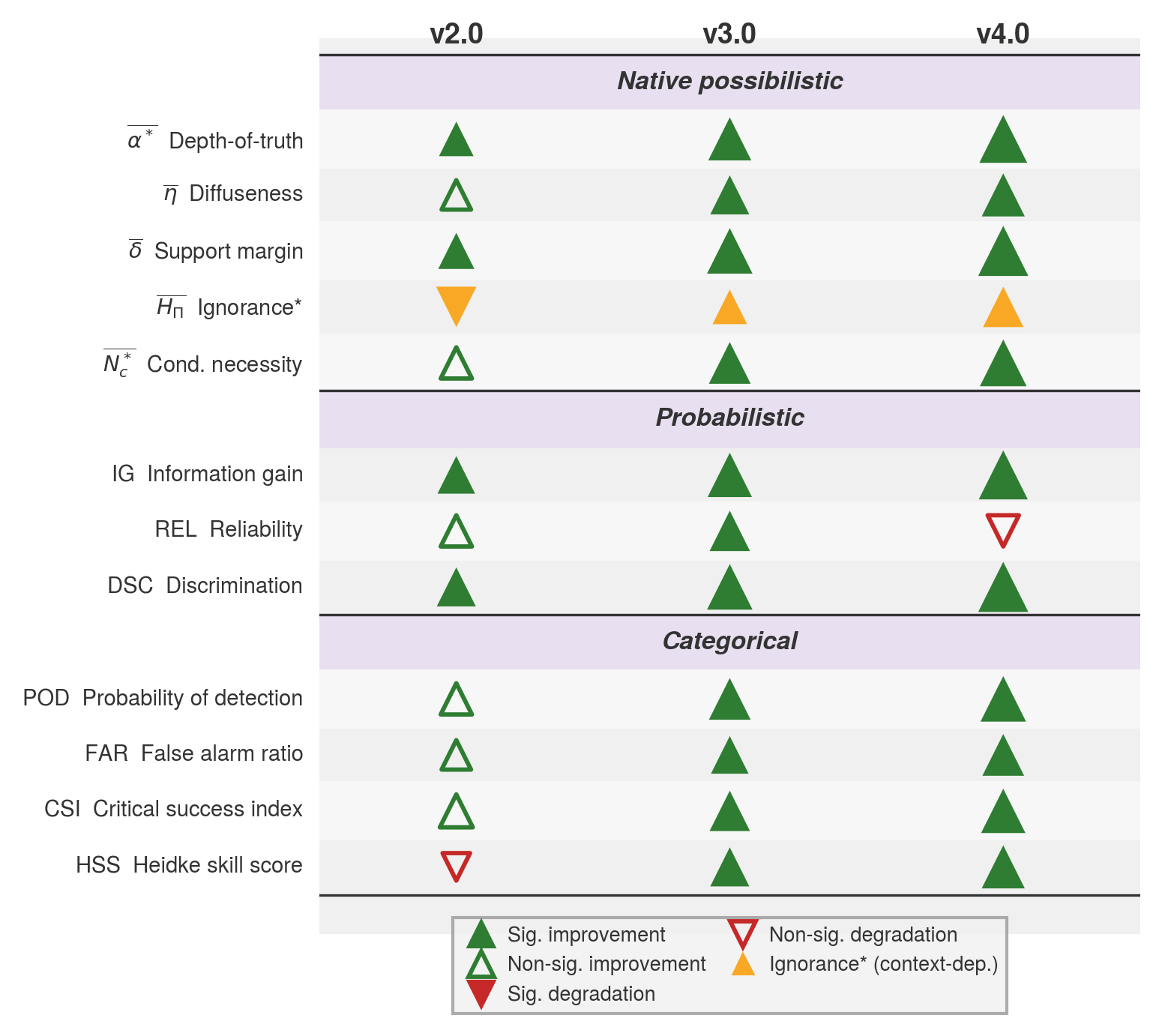}
\caption{Verification scorecard for a hypothetical possibilistic
    forecasting system across three development versions, each compared
    to a v1.0 baseline. Rows are grouped into three verification facets:
    native possibilistic (five-number scorecard), probabilistic
    (information gain and its decomposition),
    and categorical (POD, FAR, CSI, HSS). Filled triangles indicate
    statistically significant improvement (green, upward) or degradation
    (red, downward); open triangles indicate non-significant changes.
    Marker size encodes the magnitude of the change. Note the
    $\overline{H_\Pi}$ degradation in v2.0: the system became more
    ignorant before improving in subsequent versions---a signature of
    recalibrated uncertainty that the native possibilistic facet detects
    but the probabilistic facet may absorb.
    Ignorance changes (amber) are context-dependent: increased $\ign$
    may indicate honest self-awareness or declining skill, depending on
    whether discrimination also changes.}
\label{fig:scorecard_table}
\end{figure}

In practice, an evaluator would compile scorecard aggregates
across a test sample, tracking changes between model versions or
seasons wit, say, scorecards \citep{Gallo2019-nc}.
Figure~\ref{fig:scorecard_table} illustrates
this convention using a arbitrary example
for an example comparison of four model versions.
Metrics are grouped by verification type;
filled triangles denote statistically significant
change (method at researcher's discretion)
and marker size encodes magnitude,
following the scorecard style used by ECMWF
and WMO Lead Centres for NWP inter-comparisons.

The scorecard provides an accessible overview
at a glance:
a column of large green triangles means the new
version is broadly better;
a red triangle amidst green ones flags a trade-off
that warrants investigation.
The example in Fig.~\ref{fig:scorecard_table}
tells a narrative that version~2.0
improved discrimination
but increased ignorance (the system learned to hedge more honestly);
version~3.0 delivered broad gains across all three facets; version~4.0
pushed discrimination further at the cost of slight reliability
degradation.
These unavoidable trade-offs are immediately visible
and provides insights a single-metric evaluation
would likely miss.

\FloatBarrier


\section{Worked Examples}\label{sec:worked_examples}

To further aid interpretation, discussion that follows
applies the range of verification metrics given herein
to three canonical examples of convective outlooks:
sharp-correct, hedged-correct, and sharp-wrong.

\subsection{Scenario Setup}\label{sec:scenario_setup}

As introduced in Section~\ref{fig:ig_decomp},
the thought experiment is a notional possibilistic
"shadow" forecast where the actual Day-1 outlook
stands in a "verification".
Given its purely illustratory application herein,
despite the awkward concept,
this like-for-like comparison ensures both forecasts
share the same categorical universe.

\paragraph{Scenario A: Sharp-correct \spc{MDT} forecast.}

A textbook severe-weather environment: all ingredients for significant
severe thunderstorms are present and co-located. The system assigns
high possibility to \spc{MDT} with low subnormality ($m = 0.90$,
$\ign = 0.10$) --- nearly fully committed.

\[
\pi_A(\omega) = \begin{cases}
0.00 & \omega = \spc{NONE} \\
0.00 & \omega = \spc{MRGL} \\
0.05 & \omega = \spc{SLGT} \\
0.15 & \omega = \spc{ENH}  \\
0.90 & \omega = \spc{MDT}  \\
0.10 & \omega = \spc{HIGH}
\end{cases}
\]
Observed outcome: $c_{\mathrm{obs}} = \spc{MDT}$.

\paragraph{Scenario B: Hedged-correct \spc{ENH} forecast.}

Competing signals of moderate instability and adequate shear mean the
mesoscale trigger is uncertain. The system hedges across four
categories with high subnormality ($m = 0.55$, $\ign = 0.45$),
explicitly advertising limited confidence.

\[
\pi_B(\omega) = \begin{cases}
0.10 & \omega = \spc{NONE} \\
0.10 & \omega = \spc{MRGL} \\
0.40 & \omega = \spc{SLGT} \\
0.55 & \omega = \spc{ENH}  \\
0.30 & \omega = \spc{MDT}  \\
0.00 & \omega = \spc{HIGH}
\end{cases}
\]
Observed outcome: $c_{\mathrm{obs}} = \spc{ENH}$.

\paragraph{Scenario C: Sharp-wrong \spc{NONE} forecast.}

Model guidance unanimously favours no severe weather. The system
confidently assigns high possibility to \spc{NONE} and near-zero to all
severe categories. In reality, an unresolved mesoscale feature triggers
an unexpected severe outbreak that verifies at \spc{MDT}, a category
the system deemed essentially impossible.

\[
\pi_C(\omega) = \begin{cases}
0.85 & \omega = \spc{NONE} \\
0.10 & \omega = \spc{MRGL} \\
0.05 & \omega = \spc{SLGT} \\
0.00 & \omega = \spc{ENH}  \\
0.00 & \omega = \spc{MDT}  \\
0.00 & \omega = \spc{HIGH}
\end{cases}
\]
Observed outcome: $c_{\mathrm{obs}} = \spc{MDT}$.

\subsection{Categorical Example}\label{sec:categorical_walkthrough}

\begin{figure*}[t]
\centering
\includegraphics[width=0.90\textwidth]{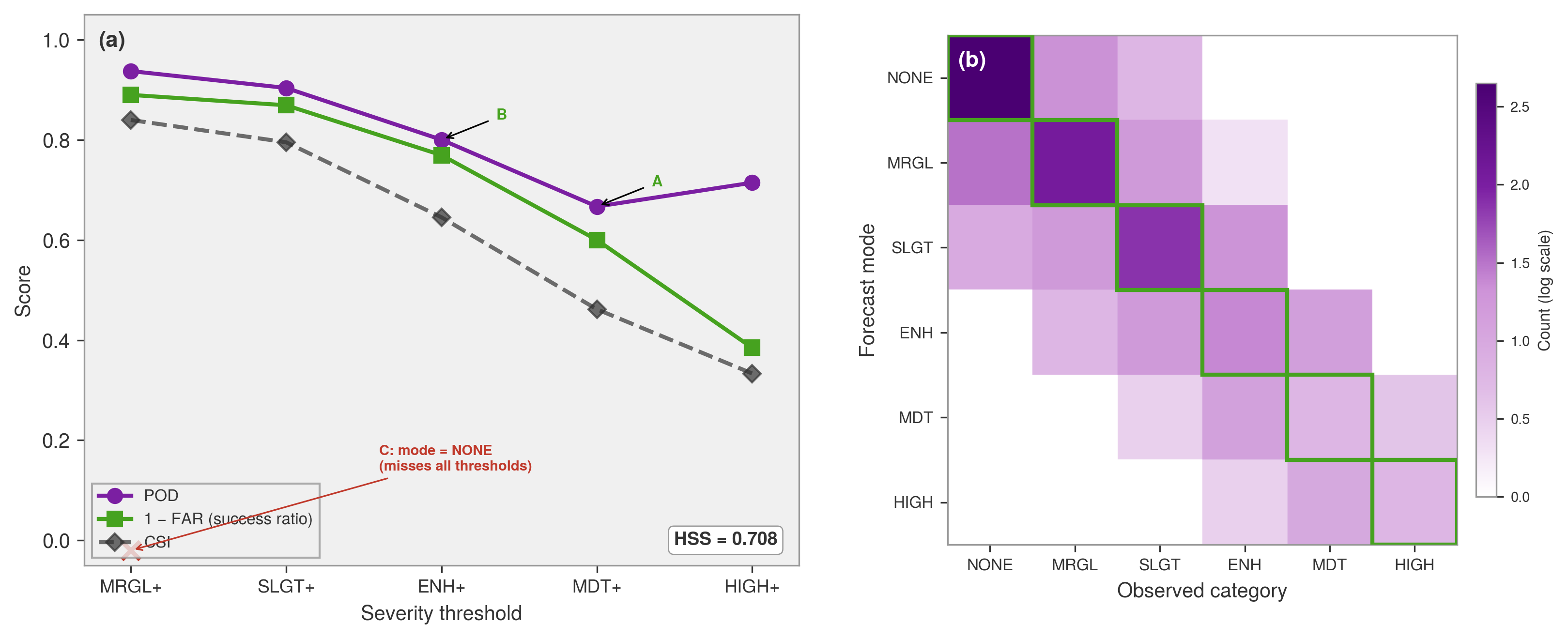}
\caption{Categorical-facet verification for the 800-day synthetic
    reforecast.
    \textbf{(a)}~Threshold performance: POD (purple), success ratio
    $1 - \text{FAR}$ (green), and CSI (dashed grey) at each severity
    threshold. Scenarios~A and~B (green labels) contribute hits at
    their respective thresholds; Scenario~C (red cross) registers
    POD~$= 0$ at all thresholds. HSS is computed from the full
    $K \times K$ table.
    \textbf{(b)}~Confusion matrix of forecast peak category vs.\ observed
    category. Green boxes mark the diagonal (correct peak); counts
    are on a log scale for visibility. Off-diagonal mass clusters
    within one category of the diagonal, indicating near-miss errors.}
\label{fig:categorical_scores}
\end{figure*}

The categorical facet reduces the full distribution to a single
deterministic category: the peak category
$\hat{c} = \arg\max_\omega \pi(\omega)$ (the system's top pick; ties
broken by severity). All other distributional information is discarded.
Verification then proceeds as a standard multi-threshold contingency
problem: for each severity threshold $t$, the forecast issues ``yes'' if
$\hat{c} \geq t$ and ``no'' otherwise; the observation is ``yes'' if
$c_{\mathrm{obs}} \geq t$.

For Scenario~A the peak category is $\hat{c} = \spc{MDT}$, matching
$c_{\mathrm{obs}} = \spc{MDT}$---a hit at every threshold up to
\spc{MDT}+. For Scenario~B the peak category is $\hat{c} = \spc{ENH}$,
again matching the observation. Scenario~C's peak category is
$\hat{c} = \spc{NONE}$ while $c_{\mathrm{obs}} = \spc{MDT}$---a miss
at every threshold, yielding POD~$= 0$ across the board.

Figure~\ref{fig:categorical_scores} places these cases in the context of
the 800-day synthetic reforecast. Panel~(a) shows how POD (purple),
success ratio $1 - \mathrm{FAR}$ (green), and CSI (dashed) degrade as
the severity threshold rises: the system detects \spc{MRGL}+ events
reliably but struggles at \spc{MDT}+ and above, where both POD and CSI
drop sharply. Scenarios~A and~B register as hits at their respective
thresholds (green labels); Scenario~C misses everywhere (red cross).
Panel~(b) shows the full $K \times K$ confusion matrix: diagonal
dominance confirms that the system's peak-category forecast is broadly accurate,
while the near-diagonal clustering of errors means misses are typically
$\pm 1$ category rather than catastrophic. The HSS of $0.708$ is
strong but not exceptional---the categorical facet alone cannot
distinguish the confident sharp-correct forecast (A) from an equally
accurate but heavily hedged one.

\subsection{Probabilistic Example}\label{sec:bridge_walkthrough}

Recall that the possibility-to-probability conversion reserves the
ignorance fraction $\ign$ as an explicit $(n{+}1)$-th outcome,
distributing the remaining mass proportionally among the original
categories (Section~\ref{sec:poss_prob_bridge}).

The conversion is now applied to each scenario. Information gain is then
computed relative to the SPC climatological baseline
$p_{\mathrm{clim}} = (0.60,\, 0.18,\, 0.12,\, 0.06,\, 0.032,\,
0.008)$, in which \spc{NONE} dominates and \spc{HIGH} is rare.
The conversion reserves $p_{\mathrm{ign}} = \ign$, distributes
the remaining $m$ proportionally via
$p_i = \pi_i \cdot m / \sum_j \pi_j$,
and appends the ignorance outcome.
Table~\ref{tab:bridge_ils} collects the results; the key pattern is
readable column by column.

\begin{table}[htbp]
\centering
\caption{Probability conversion and information gain for the three worked
    scenarios, relative to the SPC climatological baseline.
    Asterisk ($^*$) denotes $\varepsilon$-floored values
    ($\varepsilon = 0.01$; see text).}
\label{tab:bridge_ils}
\footnotesize
\setlength{\tabcolsep}{4pt}
\begin{tabular}{lccc}
\toprule
 & \textbf{A} (sharp- & \textbf{B} (hedged- & \textbf{C} (sharp- \\
 & correct) & correct) & wrong) \\
\midrule
$\ign$ & $0.10$ & $0.45$ & $0.15$ \\
$\sum \pi$ & $1.20$ & $1.45$ & $1.00$ \\
$p(c_{\mathrm{obs}})$ & $0.675$ & $0.209$ & $0.010^*$ \\
$p_{\mathrm{clim}}(c_{\mathrm{obs}})$ & $0.0320$ & $0.0600$ & $0.0320$ \\
Clim.\ surprise (bits) & $4.966$ & $4.059$ & $4.966$ \\
Forecast surprise (bits) & $0.567$ & $2.261$ & $6.644^*$ \\
\textbf{IG (bits)} & $\mathbf{+4.399}$ & $\mathbf{+1.798}$ & $\mathbf{-1.678}$ \\
\bottomrule
\end{tabular}
\end{table}

The three scenarios trace a clear arc from skill to catastrophe.
Scenario~A's derived probabilities concentrate on the correct outcome
($p = 0.675$); low ignorance means little mass is stranded in the
ignorance bin, yielding $\mathrm{IG} = +4.399$~bits---strong
skill relative to the climatological baseline, which assigns only
$3.2\%$ probability to \spc{MDT}.
Scenario~B is positive and substantial ($\mathrm{IG} = +1.798$~bits):
the high $\ign = 0.45$ deflates $p(\spc{ENH})$ from the
simple-normalisation value of $0.55/1.45 = 0.379$ down to $0.209$,
preserving the system's uncertainty, but the forecast still
beats the $6\%$ climatological rate for \spc{ENH} by a wide margin.
Scenario~C incurs a penalty ($\mathrm{IG} = -1.678$~bits) because the
conversion assigns zero probability to the observed outcome. The
$\varepsilon$-floor ($\varepsilon = 0.01$) prevents an infinite
penalty but still yields a surprise of $6.644$~bits --- worse than
the climatological surprise of $4.966$~bits for \spc{MDT}. Crucially,
low ignorance ($\ign = 0.15$) provides little cushion: the system
left itself no escape route for an outcome it had ruled impossible.

The $\varepsilon$-floor prevents the logarithmic score
from diverging ($-\log_2(0) = \infty$) when the forecast assigns zero
probability to the observed outcome. The choice $\varepsilon = 0.01$
is subjective, corresponding roughly to the quantisation floor for a
system built on $\order{100}$ ensemble members and human
post-processing. Because only $p(c_{\mathrm{obs}})$ enters the
surprise (Eq.~\ref{eq:surprise}), the full $(n{+}1)$-category
probability vector need not sum exactly to unity after flooring.
Sensitivity analysis confirms the qualitative
conclusion is invariant and is left to the
evaluator's subjectivity:

\begin{itemize}
    \item $\varepsilon = 0.001$ yields $9.97$~bits surprise,
    \item $\varepsilon = 0.0001$ yields $13.29$~bits.
\end{itemize}

While the choice of $\varepsilon$ is important to
test as above,
we find that Scenario~C remains strongly negative
no matter the enforced floor.
The author submits the subjective choice of
$\varepsilon$ is less important than the desired
behaviour of information-theoretical
scores \citep{McCutcheon2019-jq}.

\subsection{Full Scorecard Comparison}\label{sec:scorecard_comparison}

\begin{figure*}[t]
\centering
\includegraphics[width=0.90\textwidth]{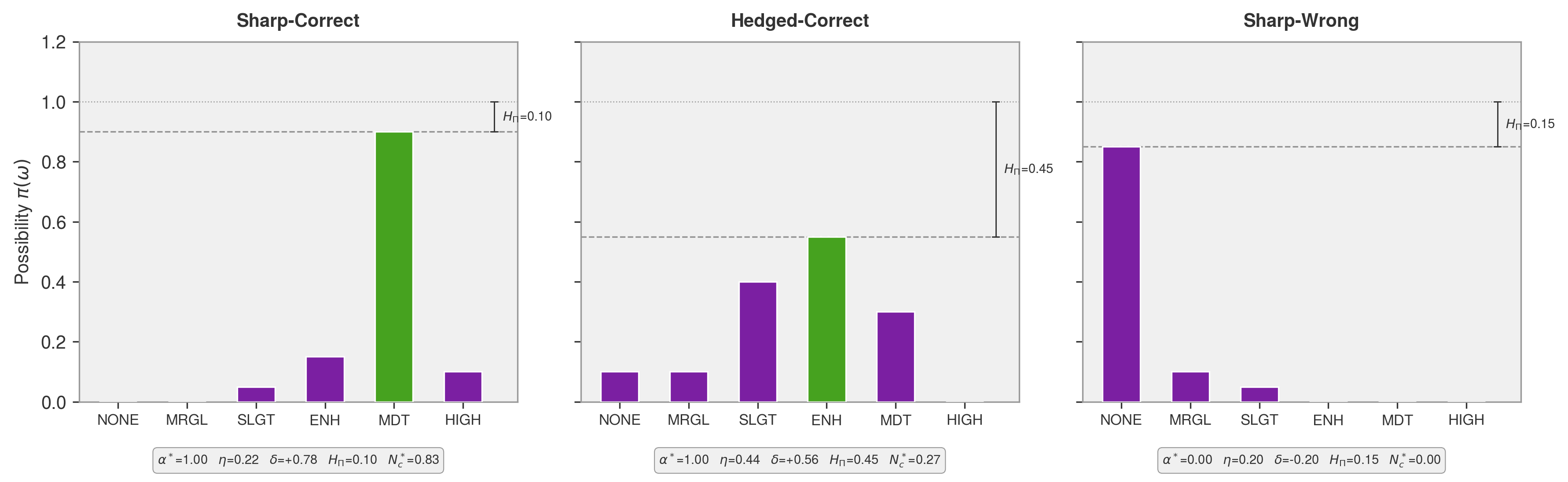}
\caption{Three-scenario comparison. Each panel shows the raw possibility
    distribution $\pi(\omega)$ as bar charts; the observed category is
    highlighted in green, others in purple. The five-number scorecard
    values are printed below each panel. \textbf{(a)}~Scenario~A:
    sharp-correct \spc{MDT}---high $\alpha^*$, high $\Nc^*$, positive
    $\delta$. \textbf{(b)}~Scenario~B: hedged-correct \spc{ENH}---high
    $\alpha^*$ but diffuse shape ($\eta$) and low $\Nc^*$.
    \textbf{(c)}~Scenario~C: sharp-wrong
    \spc{NONE}---$\alpha^* = 0$ (truth assigned zero possibility) and
    negative $\delta$.}
\label{fig:three_scenario}
\end{figure*}

Table~\ref{tab:scorecard_comparison} collects the five-number scorecard
for all three scenarios. Figure~\ref{fig:three_scenario} provides a
visual comparison, with scorecard values printed below each panel.

\begin{table}[htbp]
\centering
\caption{Five-number scorecard for the three worked scenarios.}
\label{tab:scorecard_comparison}
\footnotesize
\setlength{\tabcolsep}{4pt}
\begin{tabular}{lccc}
\toprule
 & \textbf{A} (sharp- & \textbf{B} (hedged- & \textbf{C} (sharp- \\
 & correct) & correct) & wrong) \\
\midrule
$m$ (commitment)       & $0.90$  & $0.55$  & $0.85$ \\
$\ign$ (ignorance)     & $0.10$  & $0.45$  & $0.15$ \\
$\alpha^*$ (depth-of-truth) & $1.00$ & $1.00$ & $0.00$ \\
$\eta$ (diffuseness) & $0.222$ & $0.439$ & $0.196$ \\
$\delta$ (support margin) & $+0.778$ & $+0.561$ & $-0.196$ \\
$\Nc^*$ (cond.\ necessity) & $0.833$ & $0.273$ & $0.00$ \\
\bottomrule
\end{tabular}
\end{table}

Scenario~A is the ideal case: $\alpha^* = 1$, $\Nc^* = 0.833$,
$\delta = +0.778$---the system pointed at the truth and said so
confidently ($\ign = 0.10$).

Scenario~B tells a different but equally honest story. Depth-of-truth
remains perfect ($\alpha^* = 1$) because \spc{ENH} is the possibilistic
peak, but the hedging shows up: $\ign = 0.45$, $\eta = 0.439$,
$\Nc^* = 0.273$. The support margin $\delta = +0.561$ is positive
but moderate, suggesting latent skill tempered by acknowledged uncertainty.

Scenario~C is a worst-case outcome. The system declared \spc{MDT} impossible
($\pi = 0$), so $\alpha^* = 0$ and $\Nc^* = 0$. The support margin
$\delta = -0.196$ is the only negative value in the table: the forecast
pointed \emph{away} from the truth. Critically, ignorance was low
($\ign = 0.15$)---the system was confident and wrong, with no
uncertainty buffer to absorb the miss.

The scorecard separates ``honestly uncertain'' (Scenario~B:
$\delta = +0.561$) from ``confidently wrong'' (Scenario~C:
$\delta = -0.196$)---precisely the distinction that operational
forecasters need when tuning a possibilistic system.

Given the inability to conclusively call a
``possibilistic bust" in the manner of a binary
forecast (tornado observed or not),
the reader may ask:
\textbf{why not floor everything at $\varepsilon$?}
From scenario~C, $\alpha^* = 0$ invites this objection: why would a
system declare a category impossible when a small floor
($\pi \geq \varepsilon$) would avoid the zero? Flooring does help on
misses---Scenario~C's $\delta$ improves by $+0.006$---but it inflates
$\eta$ on \textit{every} forecast, costing Scenario~A $-0.004$ in
$\delta$. Because correct forecasts vastly outnumber catastrophic
misses in any operational sample, the net effect of flooring is
negative: the chronic diffuseness tax outweighs the rare insurance
payout. Declaring a category impossible is valuable claim
when correct---it drives $\eta$ down and $\delta$ up across the
verification sample---but as the counterpart,
the scorecard penalises overconfidence
appropriately harshly when wrong.

The reader may also ask for more clarity:
\textbf{why five numbers?} Three diagnostic components
(Table~\ref{tab:three_to_five}) expand to five post-event metrics
because the possibility component requires three numbers to capture
support for the truth ($\alpha^*$), forecast spread ($\eta$), and their
net balance ($\delta$).
While the scorecard is constructed from insightful
metrics subjectively,
Section~\ref{sec:tripartite_value} provides the
formal justification;
the three scenarios make the complementarity
tangible:
\begin{itemize}[leftmargin=2em]
\item \textbf{Truth-finding pair} ($\alpha^*$ vs.\ $\Nc^*$).
    Scenario~B has $\alpha^* = 1.00$ but $\Nc^* = 0.273$: truth was the
    top pick but closely contested by alternatives. Had the observed
    category been \spc{ENH} instead of \spc{MDT} in Scenario~A's
    distribution, $\alpha^*$ would drop to $0.167$ while $\Nc^*$ would
    remain zero---truth was plausible but far from dominant.
    $\alpha^*$ answers ``how much support did the truth receive?'';
    $\Nc^*$ answers ``was the truth \textit{the} winner?''
\item \textbf{Uncertainty pair} ($\eta$ vs.\ $\ign$).
    Scenario~B has $\eta = 0.439$ and $\ign = 0.45$: the system was both
    hedged in shape and uncertain in commitment. Scenario~C has
    $\eta = 0.196$ and $\ign = 0.15$: sharp in shape \textit{and}
    confident in commitment---but aimed at the wrong answer.
    $\eta$ measures shape spread (normalised); $\ign$ measures overall
    confidence (raw). They move independently.
\item \textbf{Net support margin} ($\delta$).
    The one number that combines truth-finding with spread: $\delta > 0$
    means the forecast shape helped; $\delta < 0$ conversely so.
    No other single scorecard metric captures this balance.
\end{itemize}

Indeed, a reduced scorecard might conflate diagnostically distinct situations:
dropping $\Nc^*$ loses the ``winner vs.\ supported'' distinction;
dropping $\ign$ hides the confidence axis; dropping $\delta$ obscures
whether the shape pointed toward or away from the truth.


\section{Discussion and Future Work}\label{sec:discussion}

The author herein developed a verification framework
for possibilistic forecasts
(i.e., uncertain probabilities) over finite
forecast categories.
The core contribution is a five-number
scorecard (Section~\ref{sec:native_scorecard}) that evaluates a
subnormal (i.e., ignorance explicitly measured)
possibility distribution on its own terms, without first
converting it to a probability. Supporting this scorecard are a
possibility-to-probability transformation
(Section~\ref{sec:poss_prob_bridge}) that preserves the ignorance
signal as an explicit outcome rather than erasing it through
normalisation, and three complementary verification facets
(Section~\ref{sec:three_lanes}) that together cover
categorical, probabilistic, and possibilistic evaluation.
Worked examples in Section~\ref{sec:worked_examples} illustrated
that each uncertainty component---possibility, ignorance, and
conditional necessity---can contribute distinct verification value,
so that collapsing the distribution into a single point forecast
discards actionable information. Although the SPC convective outlook
categories served as the running example throughout, the framework
applies to any finite universe of discourse $\Omega$ over which a
subnormal possibility distribution is defined.

\subsection{Generating Possibilistic Forecasts}%
\label{sec:generation}

A natural question arises: where do possibilistic forecasts come from?
This paper addresses verification, not generation, but several pathways
exist for producing the subnormal distributions the framework evaluates.

\textbf{Fuzzy inference systems.}  The most direct route uses fuzzy
rules whose membership functions are possibility distributions by
construction \citep{Klir1995-va}. A fuzzy inference system maps input
meteorological variables through rulesets with fuzzy
categories to an output forecast.
Operational examples exist
(e.g., \citealt{Asklany2011-yn,Camastra2015-by}),
including Bingham Research Center's possibilistic
forecasts for Utah air
quality \citep{Lawson2024-jb}.

\textbf{Ensemble-derived possibility.}
The UK Met Office continue to improve production of
severity--confidence matrices from ensemble
forecasts \citep{Taylor2024-pz,Neal2014-jg,Taylor2019-is}.
Framing ensemble spread in terms of subnormality
may be the most tangible pathway for an NWP
audience, and post-processing provides the
opportunity to deploy methods capable of capturing
edge cases with low
confidence \citep{Williams2014-jb}.
Increased nuance applied to National Weather
Service (NWS)
warnings \citep{Trujillo-Falcon2022-du} continue
to extract richer warning guidance tailored to
local population.

\begin{figure}[t]
\centering
\includegraphics[width=\columnwidth]{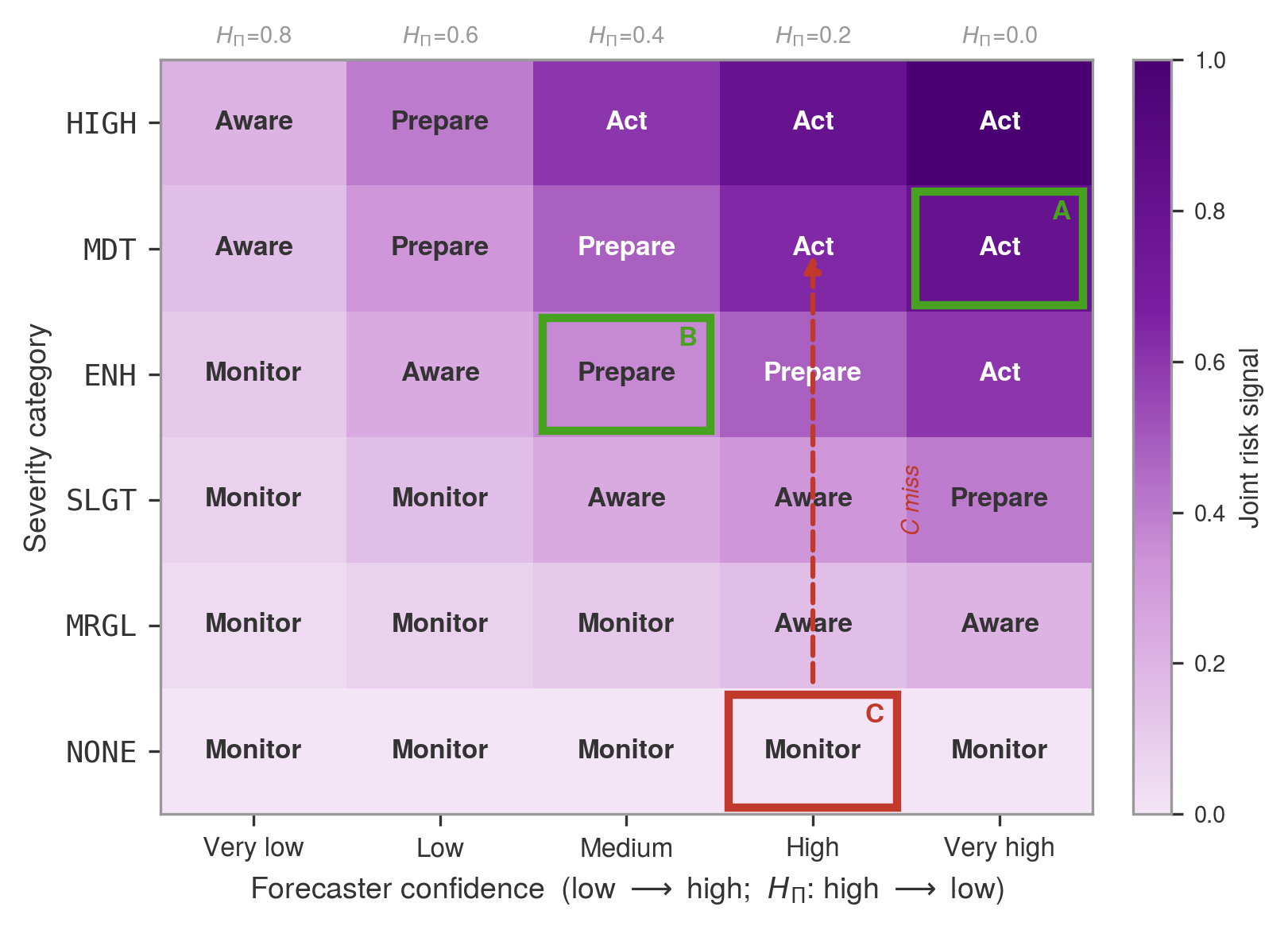}
\caption{Severity--confidence matrix adapted from the UK Met Office
    NSWWS framework \citep{Neal2014-jg} into the SPC categorical
    system. Rows are SPC severity categories (\spc{NONE}--\spc{HIGH});
    columns represent forecaster confidence, mapped to subnormality
    ($\ign$). Colour intensity encodes the joint risk signal: deep purple
    indicates high severity with high confidence (act now); pale tones
    indicate low severity or low confidence (monitor). Worked scenarios from
    Section~\ref{sec:worked_examples} are placed in the nearest
    confidence column (discrete binning); exact $\ign$ values may fall
    between column centres.}
\label{fig:severity_matrix}
\end{figure}

\textbf{Severity--confidence matrices.}  Operational warning frameworks
already encode possibilistic structure. The UK Met Office National
Severe Weather Warning Service (NSWWS) uses a two-dimensional matrix
of likelihood and impact to determine warning levels
\citep{Neal2014-jg}. The impact axis encodes what \textit{might}
happen (the shape of the possibility distribution), while the likelihood
axis encodes forecast confidence (commitment or subnormality).
Figure~\ref{fig:severity_matrix} adapts this concept to the paper's SPC
framework for illustration purposes,
mapping severity categories against confidence
levels to show
how the joint signal varies from ``monitor'' (low severity, high
ignorance) to ``act now'' (high severity, low ignorance). The
verification framework developed is a mere exercise
and not a serious proposal for risk communication
today \citep{Rothfusz2018-yk};
it is deployed only to formalise an operational
intuition with precise metrics.

To connect the matrix to the verification framework, consider a cell at
\spc{MDT} severity with high confidence ($\ign = 0.10$). The implied
distribution might be
$\pi = (0.00,\, 0.00,\, 0.05,\, 0.15,\, 0.90,\, 0.10)$---exactly
Scenario~A from Section~\ref{sec:worked_examples}. The five-number
scorecard then evaluates whether that matrix cell's implied distribution
is well-calibrated: $\alpha^* = 1.0$ and $\delta = +0.778$ when
\spc{MDT} verifies, but $\alpha^* = 0$ and $\delta < 0$ when it does
not. Aggregating scorecard values across all days that fall in a given
matrix cell add insight to whether the operational severity--confidence mapping
is empirically justified.


\subsection{Limitations}\label{sec:limitations}

Open questions remain, including:

\begin{itemize}
    \item Creation of a possibilistic skill score;
    \item Derivation of a \textit{proper} evaluation metric
    \item How do you evaluate ignorance?
    The probability conversion assumes a closed
    world, i.e., the observed event is always
    in $\Omega$.
    Open-world applications would require the
    ignorance outcome to verify as ``correct.''
\end{itemize}

In earth sciences, predicted quantities are often
continuous. Herein, the scorecard treats $\Omega$
as unordered: missing by one category
(\spc{ENH} when \spc{MDT} verifies) incurs the same $\alpha^*$ penalty
as missing by four (\spc{NONE} when \spc{HIGH} verifies). This suits
domains without inherent ordering (e.g., precipitation type) but
discards near-miss information for ordinal domains like SPC severity.
A distance-sensitive score could supplement the scorecard for such
domains. Diffuseness $\eta$ partially captures ordinal hedging, but the
scorecard does not formally consider category
adjacency (find discussion of statistical
locality in REFS).

\subsection{Future work: large-language model
(LLM) outlooks}\label{sec:future_directions}

Given the complexity and comprehensive covered
provided by the scorecard,
we must seek a complexity dial to deploy language
as plain and familiar as
necessary \citep{Gigerenzer2005-zs,Joslyn2009-rg}.
The most immediate application is risk
communication via large language models (LLMs).
The possibilistic triplet ($\poss$, $\ign$, $\Nc$) is
a compact, pre-digested summary of forecast state
that an LLM can process to yield (theoretically)
richer uncertainty discussion outlooks.
The stepping-down of a scorecard to natural language
(prose) is key for public communication of,
e.g., SPC forecasts \citep{Krocak2022-sy};
other stakeholders benefit from a full range of
forecast variables \citep{McGovern2017-kf,Le_Carrer2021-ke},
with prose at complexity sufficient for the
user's needs.
By constraining the LLM's input to structured
scorecard
values rather than dense numerical fields, the framework reduces the
context the model must hold and grounds its generation in
verifiable quantities.
Preliminary work shows large reduction in
hallucination \citep{Jardine1997-tn} when tightly
coupled with a (digestible) dataset,
tackling a similar problem
as \citet{Chen2023-ax} with a simpler method
than \citep{Jin2023-cv}.

Translating these triplets into tiered plain-language advisories is the
subject of ongoing work (Lawson, in prep.).
Precedence of using statistics to reduce
overwhelming datasets into manageable concepts or
clusters exist in large-membership NWP ensembles:
it is impossible for a human forecaster to
synthesise 50 model forecasts simultaneously;
as such, dimension reduction (thresholding;
clustering; etc) is used to lower the forecaster
burden on contextualising the situation and
forecast problem.


\section*{Acknowledgments}\label{sec:acknowledgments}
The author is funded by Uintah County Special
Service District 1 and the Utah Legislature.


\section*{Generative AI Statement}\label{sec:genai_statement}

Generative AI tools (Google Gemini, Anthropic Claude,
OpenAI Codex) were used in project brainstorming,
manuscript preparation, and code development.
The project history from inception to submission
is versioned at the GitHub repository linked below.

\section*{Data and Code Availability}\label{sec:data_code}

All verification metrics, the probability conversion, and the visualisation
tools described in this paper are implemented in Python and publicly
available at
\url{https://github.com/bingham-research-center/possibility-verif}.



\appendix



\bibliographystyle{ametsocV6}
\bibliography{paperpile}

@ARTICLE{Asklany2011-yn,
  title   = "Rainfall events prediction using rule-based fuzzy inference system",
  author  = "Asklany, Somia A and Elhelow, Khaled and Youssef, I K and Abd
             El-wahab, M",
  journal = "Atmos. Res.",
  volume  =  101,
  number  =  1,
  pages   = "228--236",
  month   =  jul,
  year    =  2011,
  url     = "https://www.sciencedirect.com/science/article/pii/S0169809511000640",
  doi     = "10.1016/j.atmosres.2011.02.015",
  issn    = "0169-8095"
}

@ARTICLE{Benedetti2010-sa,
  title     = "Scoring Rules for Forecast Verification",
  author    = "Benedetti, Riccardo",
  journal   = "Mon. Weather Rev.",
  publisher = "American Meteorological Society",
  volume    =  138,
  number    =  1,
  pages     = "203--211",
  month     =  jan,
  year      =  2010,
  url       = "https://doi.org/10.1175/2009MWR2945.1",
  doi       = "10.1175/2009MWR2945.1",
  issn      = "0027-0644"
}

@ARTICLE{Brier1950-lc,
  title     = "Verification Of Forecasts Expressed In Terms Of Probability",
  author    = "Brier, Glenn W",
  journal   = "Mon. Weather Rev.",
  publisher = "American Meteorological Society",
  volume    =  78,
  number    =  1,
  pages     = "1--3",
  month     =  jan,
  year      =  1950,
  url       = "http://journals.ametsoc.org/doi/abs/10.1175/1520-0493(1950)078%3C0001:VOFEIT%3E2.0.CO;2",
  eprint    = "http://dx.doi.org/10.1175/1520-0493(1950)078<0001:VOFEIT>2.0.CO;2",
  doi       = "10.1175/1520-0493(1950)078<0001:VOFEIT>2.0.CO;2",
  issn      = "0027-0644,1520-0493",
  language  = "en"
}

@ARTICLE{Camastra2015-by,
  title     = "A fuzzy decision system for genetically modified plant
               environmental risk assessment using Mamdani inference",
  author    = "Camastra, Francesco and Ciaramella, Angelo and Giovannelli,
               Valeria and Lener, Matteo and Rastelli, Valentina and Staiano,
               Antonino and Staiano, Giovanni and Starace, Alfredo",
  journal   = "Expert Syst. Appl.",
  publisher = "Elsevier BV",
  volume    =  42,
  number    =  3,
  pages     = "1710--1716",
  month     =  feb,
  year      =  2015,
  url       = "http://dx.doi.org/10.1016/j.eswa.2014.09.041",
  doi       = "10.1016/j.eswa.2014.09.041",
  issn      = "0957-4174,1873-6793"
}

@ARTICLE{Chen2023-ax,
  title       = "{MisCaltral}: Reducing numeric hallucinations of Mistral with
                 precision numeric calculation",
  author      = "Chen, Shih-Wen and Hsu, Hsien-Jung",
  journal     = "Research Square",
  institution = "Research Square",
  month       =  dec,
  year        =  2023,
  url         = "https://www.researchsquare.com/article/rs-3789011/latest",
  doi         = "10.21203/rs.3.rs-3789011/v1"
}

@ARTICLE{Coleman2024-hf,
  title     = "Can we predict the predictability of high-impact weather events?",
  author    = "Coleman, Austin A and Ancell, Brian C and Schwartz, Craig S",
  journal   = "Mon. Weather Rev.",
  publisher = "American Meteorological Society",
  month     =  sep,
  year      =  2024,
  url       = "https://journals.ametsoc.org/downloadpdf/view/journals/mwre/aop/MWR-D-23-0293.1/MWR-D-23-0293.1.pdf",
  doi       = "10.1175/mwr-d-23-0293.1",
  issn      = "1520-0493,0027-0644",
  language  = "en"
}

@BOOK{Cover2012-di,
  title     = "Elements of Information Theory",
  author    = "Cover, Thomas M and Thomas, Joy A",
  publisher = "John Wiley \& Sons",
  address   = "Hoboken, NJ, USA",
  month     =  nov,
  year      =  2012,
  url       = "http://dx.doi.org/10.1002/047174882X",
  doi       = "10.1002/047174882X",
  isbn      =  9781118585771,
  language  = "en"
}

@BOOK{Dubois1988-nh,
  title     = "Possibility theory: An approach to computerized processing of
               uncertainty",
  author    = "Dubois, Didier and Prade, Henri",
  publisher = "Plenum Press",
  address   = "New York; London",
  year      =  1988,
  url       = "https://books.google.com/books?hl=en&lr=&id=GEvUBwAAQBAJ&oi=fnd&pg=PA1&ots=wzFIjIE942&sig=XmL0E58X-e4SI5UZF8CNQ2hrSog",
  isbn      =  9781468452877,
  language  = "en"
}

@ARTICLE{Dubois1992-gd,
  title     = "When upper probabilities are possibility measures",
  author    = "Dubois, Didier and Prade, Henri",
  journal   = "Fuzzy Sets And Systems",
  publisher = "Elsevier BV",
  volume    =  49,
  number    =  1,
  pages     = "65--74",
  month     =  jul,
  year      =  1992,
  url       = "http://dx.doi.org/10.1016/0165-0114(92)90110-P",
  doi       = "10.1016/0165-0114(92)90110-p",
  issn      = "0165-0114,1872-6801",
  language  = "en"
}

@ARTICLE{Dubois2007-dd,
  title     = "Comparing probability measures using possibility theory: A notion
               of relative peakedness",
  author    = "Dubois, Didier and Hüllermeier, Eyke",
  journal   = "Int. J. Approx. Reason.",
  publisher = "Elsevier BV",
  volume    =  45,
  number    =  2,
  pages     = "364--385",
  month     =  jul,
  year      =  2007,
  url       = "http://dx.doi.org/10.1016/j.ijar.2006.06.017",
  doi       = "10.1016/j.ijar.2006.06.017",
  issn      = "0888-613X,1873-4731",
  language  = "en"
}

@ARTICLE{Gallo2019-nc,
  title     = "Initial Development and Testing of a Convection-Allowing Model
               Scorecard",
  author    = "Gallo, Burkely T and Kalb, Christina P and Gotway, John Halley
               and Fisher, Henry H and Roberts, Brett and Jirak, Israel L and
               Clark, Adam J and Alexander, Curtis and Jensen, Tara L",
  journal   = "Bull. Am. Meteorol. Soc.",
  publisher = "American Meteorological Society",
  volume    =  100,
  number    =  12,
  pages     = "ES367--ES384",
  month     =  dec,
  year      =  2019,
  url       = "https://journals.ametsoc.org/view/journals/bams/100/12/bams-d-18-0218.1.xml",
  doi       = "10.1175/BAMS-D-18-0218.1",
  issn      = "0003-0007,1520-0477",
  language  = "en"
}

@ARTICLE{Gigerenzer2005-zs,
  title   = "``{A} 30\% chance of rain tomorrow'': how does the public
             understand probabilistic weather forecasts?",
  author  = "Gigerenzer, Gerd and Hertwig, Ralph and van den Broek, Eva and
             Fasolo, Barbara and Katsikopoulos, Konstantinos V",
  journal = "Risk Anal.",
  volume  =  25,
  number  =  3,
  pages   = "623--629",
  month   =  jun,
  year    =  2005,
  url     = "http://dx.doi.org/10.1111/j.1539-6924.2005.00608.x",
  doi     = "10.1111/j.1539-6924.2005.00608.x",
  pmid    =  16022695,
  issn    = "0272-4332"
}

@ARTICLE{Gneiting2007-ob,
  title     = "Strictly Proper Scoring Rules, Prediction, and Estimation",
  author    = "Gneiting, Tilmann and Raftery, Adrian E",
  journal   = "J. Am. Stat. Assoc.",
  publisher = "Taylor \& Francis",
  volume    =  102,
  number    =  477,
  pages     = "359--378",
  month     =  mar,
  year      =  2007,
  url       = "https://doi.org/10.1198/016214506000001437",
  doi       = "10.1198/016214506000001437",
  issn      = "0162-1459"
}

@MISC{Good1952-st,
  title   = "Rational Decisions",
  author  = "Good, I J",
  journal = "Journal of the Royal Statistical Society: Series B (Methodological)",
  volume  =  14,
  number  =  1,
  pages   = "107--114",
  year    =  1952,
  url     = "http://dx.doi.org/10.1111/j.2517-6161.1952.tb00104.x",
  doi     = "10.1111/j.2517-6161.1952.tb00104.x"
}

@BOOK{Green1966-xl,
  title     = "Signal detection theory and psychophysics",
  author    = "Green, David Marvin and Swets, John A and {Others}",
  publisher = "Wiley New York",
  volume    =  1,
  year      =  1966,
  url       = "http://andrei.gorea.free.fr/Teaching_fichiers/SDT%20and%20Psytchophysics.pdf"
}

@ARTICLE{Hagedorn2009-pc,
  title     = "Communicating the value of probabilistic forecasts with weather
               roulette",
  author    = "Hagedorn, Renate and Smith, Leonard A",
  journal   = "Meteorol. Appl.",
  publisher = "Wiley",
  volume    =  16,
  number    =  2,
  pages     = "143--155",
  month     =  jun,
  year      =  2009,
  url       = "https://onlinelibrary.wiley.com/doi/10.1002/met.92",
  doi       = "10.1002/met.92",
  issn      = "1350-4827,1469-8080",
  language  = "en"
}

@ARTICLE{Hendrickson1971-lx,
  title     = "Proper Scores for Probability Forecasters",
  author    = "Hendrickson, Arlo D and Buehler, Robert J",
  journal   = "aoms",
  publisher = "Institute of Mathematical Statistics",
  volume    =  42,
  number    =  6,
  pages     = "1916--1921",
  month     =  dec,
  year      =  1971,
  url       = "https://projecteuclid.org/journals/annals-of-mathematical-statistics/volume-42/issue-6/Proper-Scores-for-Probability-Forecasters/10.1214/aoms/1177693057.short",
  doi       = "10.1214/aoms/1177693057",
  issn      = "0003-4851,2168-8990",
  language  = "en"
}

@ARTICLE{Hersbach2000-yb,
  title     = "Decomposition of the Continuous Ranked Probability Score for
               Ensemble Prediction Systems",
  author    = "Hersbach, Hans",
  journal   = "Weather Forecast.",
  publisher = "American Meteorological Society",
  volume    =  15,
  number    =  5,
  pages     = "559--570",
  month     =  oct,
  year      =  2000,
  url       = "http://dx.doi.org/10.1175/1520-0434(2000)015<0559:DOTCRP>2.0.CO;2",
  doi       = "10.1175/1520-0434(2000)015<0559:DOTCRP>2.0.CO;2",
  issn      = "0882-8156"
}

@ARTICLE{Jardine1997-tn,
  title     = "Mixed messages in risk communication",
  author    = "Jardine, Cynthia G and Hrudey, Steve E",
  journal   = "Risk Anal.",
  publisher = "Wiley",
  volume    =  17,
  number    =  4,
  pages     = "489--498",
  month     =  aug,
  year      =  1997,
  url       = "https://onlinelibrary.wiley.com/doi/abs/10.1111/j.1539-6924.1997.tb00889.x",
  doi       = "10.1111/j.1539-6924.1997.tb00889.x",
  issn      = "1539-6924,0272-4332",
  language  = "en"
}

@ARTICLE{Jin2023-cv,
  title         = "Time-{LLM}: Time series forecasting by reprogramming large
                   language models",
  author        = "Jin, Ming and Wang, Shiyu and Ma, Lintao and Chu, Zhixuan and
                   Zhang, James Y and Shi, Xiaoming and Chen, Pin-Yu and Liang,
                   Yuxuan and Li, Yuan-Fang and Pan, Shirui and Wen, Qingsong",
  journal       = "arXiv [cs.LG]",
  month         =  oct,
  year          =  2023,
  url           = "http://arxiv.org/abs/2310.01728",
  archivePrefix = "arXiv",
  primaryClass  = "cs.LG",
  eprint        = "2310.01728",
  doi           = "10.48550/arXiv.2310.01728"
}

@BOOK{Jolliffe2003-xs,
  title     = "Forecast Verification: A Practitioner's Guide in Atmospheric
               Science",
  author    = "Jolliffe, Ian T and Stephenson, David B",
  publisher = "John Wiley \& Sons",
  month     =  aug,
  year      =  2003,
  url       = "https://market.android.com/details?id=book-cjS9kK8IWbwC",
  isbn      =  9780470864418,
  language  = "en"
}

@ARTICLE{Joslyn2009-rg,
  title     = "Probability of Precipitation: Assessment and Enhancement of
               End-User Understanding",
  author    = "Joslyn, Susan and {Nadav-Greenberg} and Nichols, Rebecca M",
  journal   = "Bull. Am. Meteorol. Soc.",
  publisher = "American Meteorological Society",
  volume    =  90,
  number    =  2,
  pages     = "185--194",
  month     =  feb,
  year      =  2009,
  url       = "http://dx.doi.org/10.1175/2008BAMS2509.1",
  doi       = "10.1175/2008BAMS2509.1",
  issn      = "0003-0007"
}

@ARTICLE{Klir1995-va,
  title     = "Principles of uncertainty: What are they? Why do we need them?",
  author    = "Klir, George J",
  journal   = "Fuzzy Sets And Systems",
  publisher = "Elsevier BV",
  volume    =  74,
  number    =  1,
  pages     = "15--31",
  month     =  aug,
  year      =  1995,
  url       = "https://www.sciencedirect.com/science/article/pii/016501149500032G",
  doi       = "10.1016/0165-0114(95)00032-g",
  issn      = "0165-0114,1872-6801",
  language  = "en"
}

@ARTICLE{Krocak2022-sy,
  title     = "Exploring the differences in {SPC} convective outlook
               interpretation using categorical and numeric information",
  author    = "Krocak, Makenzie J and Ripberger, Joseph T and Ernst, Sean and
               Silva, Carol L and Jenkins-Smith, Hank C",
  journal   = "Weather Forecast.",
  publisher = "American Meteorological Society",
  volume    =  37,
  number    =  2,
  pages     = "303--311",
  month     =  feb,
  year      =  2022,
  url       = "http://dx.doi.org/10.1175/WAF-D-21-0123.1",
  doi       = "10.1175/waf-d-21-0123.1",
  issn      = "0882-8156,1520-0434"
}

@ARTICLE{Lawson2024-bu,
  title     = "Decoding the atmosphere: Optimising probabilistic forecasts with
               information gain",
  author    = "Lawson, John R and Potvin, Corey K and Nelson, Kenric",
  journal   = "Meteorology",
  publisher = "MDPI AG",
  volume    =  3,
  number    =  2,
  pages     = "212--231",
  month     =  apr,
  year      =  2024,
  url       = "https://www.mdpi.com/2674-0494/3/2/10",
  doi       = "10.3390/meteorology3020010",
  issn      = "2674-0494",
  language  = "en"
}

@ARTICLE{Lawson2024-jb,
  title     = "A preliminary fuzzy inference system for predicting atmospheric
               ozone in an intermountain basin",
  author    = "Lawson, John R and Lyman, Seth N",
  journal   = "Air",
  publisher = "MDPI AG",
  volume    =  2,
  number    =  3,
  pages     = "337--361",
  month     =  sep,
  year      =  2024,
  url       = "https://www.mdpi.com/2813-4168/2/3/20",
  doi       = "10.3390/air2030020",
  issn      = "2813-4168,2813-4168",
  language  = "en"
}

@ARTICLE{Lawson2024-jc,
  title         = "Communicating risk with possibility, not probability",
  author        = "Lawson, John R",
  journal       = "arXiv [stat.AP]",
  month         =  oct,
  year          =  2024,
  url           = "https://doi.org/10.48550/arXiv.2410.21664",
  archivePrefix = "arXiv",
  eprint        = "2410.21664",
  doi           = "10.48550/arXiv.2410.21664",
  eprintclass   = "stat.AP"
}

@ARTICLE{Le_Carrer2021-by,
  title     = "Beyond probabilities: A possibilistic framework to interpret
               ensemble predictions and fuse imperfect sources of information",
  author    = "Le Carrer, Noémie and Ferson, Scott",
  journal   = "Q. J. R. Meteorol. Soc.",
  publisher = "Wiley",
  volume    =  147,
  number    =  739,
  pages     = "3410--3433",
  month     =  jul,
  year      =  2021,
  url       = "http://dx.doi.org/10.1002/qj.4135",
  doi       = "10.1002/qj.4135",
  issn      = "0035-9009,1477-870X",
  language  = "en"
}

@ARTICLE{Le_Carrer2021-ke,
  title     = "Possibly extreme, probably not: Is possibility theory the route
               for risk‐averse decision‐making?",
  author    = "Le Carrer, Noémie",
  journal   = "Atmos. Sci. Lett.",
  publisher = "Wiley",
  volume    =  22,
  number    =  7,
  month     =  jul,
  year      =  2021,
  url       = "http://dx.doi.org/10.1002/asl.1030",
  doi       = "10.1002/asl.1030",
  issn      = "1530-261X",
  language  = "en"
}

@ARTICLE{McCutcheon2019-jq,
  title     = "In Favor of Logarithmic Scoring",
  author    = "McCutcheon, Randall G",
  journal   = "Philos. Sci.",
  publisher = "Cambridge University Press",
  volume    =  86,
  number    =  2,
  pages     = "286--303",
  month     =  apr,
  year      =  2019,
  url       = "https://www.cambridge.org/core/journals/philosophy-of-science/article/abs/in-favor-of-logarithmic-scoring/72B63BA2CE6720469D23F6AD5F8F9544",
  doi       = "10.1086/702028",
  issn      = "0031-8248,1539-767X"
}

@ARTICLE{McGovern2017-kf,
  title     = "Using Artificial Intelligence to Improve Real-Time Decision
               Making for High-Impact Weather",
  author    = "McGovern, Amy and Elmore, Kimberly L and Gagne, David John and
               Haupt, Sue Ellen and Karstens, Christopher D and Lagerquist, Ryan
               and Smith, Travis and Williams, John K",
  journal   = "Bull. Am. Meteorol. Soc.",
  publisher = "American Meteorological Society",
  month     =  mar,
  year      =  2017,
  url       = "http://dx.doi.org/10.1175/BAMS-D-16-0123.1",
  doi       = "10.1175/BAMS-D-16-0123.1",
  issn      = "0003-0007"
}

@ARTICLE{Murphy1984-ox,
  title     = "Probability Forecasting in Meteorology",
  author    = "Murphy, Allan H and Winkler, Robert L",
  journal   = "J. Am. Stat. Assoc.",
  publisher = "Taylor \& Francis",
  volume    =  79,
  number    =  387,
  pages     = "489--500",
  month     =  sep,
  year      =  1984,
  url       = "https://doi.org/10.1080/01621459.1984.10478075",
  doi       = "10.1080/01621459.1984.10478075",
  issn      = "0162-1459"
}

@ARTICLE{Neal2014-jg,
  title     = "Ensemble based first guess support towards a risk-based severe
               weather warning service: Ensemble severe weather forecasts",
  author    = "Neal, Robert A and Boyle, Patricia and Grahame, Nicholas and
               Mylne, Kenneth and Sharpe, Michael",
  journal   = "Meteorol. Appl.",
  publisher = "Wiley",
  volume    =  21,
  number    =  3,
  pages     = "563--577",
  month     =  jul,
  year      =  2014,
  url       = "http://dx.doi.org/10.1002/met.1377",
  doi       = "10.1002/met.1377",
  issn      = "1469-8080,1350-4827",
  language  = "en"
}

@ARTICLE{Oussalah2002-os,
  title     = "On the normalization of subnormal possibility distributions: New
               investigations",
  author    = "Oussalah, Mourad",
  journal   = "Int. J. Gen. Syst.",
  publisher = "Taylor \& Francis",
  volume    =  31,
  number    =  3,
  pages     = "277--301",
  month     =  may,
  year      =  2002,
  url       = "https://doi.org/10.1080/03081070290005203",
  doi       = "10.1080/03081070290005203",
  issn      = "0308-1079"
}

@ARTICLE{Peirolo2011-sl,
  title   = "Information gain as a score for probabilistic forecasts",
  author  = "Peirolo, Riccardo",
  journal = "Met. Apps",
  volume  =  18,
  number  =  1,
  pages   = "9--17",
  month   =  mar,
  year    =  2011,
  url     = "http://doi.wiley.com/10.1002/met.188",
  doi     = "10.1002/met.188",
  issn    = "1350-4827"
}

@ARTICLE{Roebber2009-rv,
  title     = "Visualizing Multiple Measures of Forecast Quality",
  author    = "Roebber, Paul J",
  journal   = "Weather Forecast.",
  publisher = "American Meteorological Society",
  volume    =  24,
  number    =  2,
  pages     = "601--608",
  month     =  apr,
  year      =  2009,
  url       = "https://doi.org/10.1175/2008WAF2222159.1",
  doi       = "10.1175/2008WAF2222159.1",
  issn      = "0882-8156"
}

@ARTICLE{Rothfusz2018-yk,
  title     = "{FACETs}: A Proposed Next-Generation Paradigm for High-Impact
               Weather Forecasting",
  author    = "Rothfusz, Lans P and Schneider, Russell and Novak, David and
               Klockow-McClain, Kimberly and Gerard, Alan E and Karstens, Chris
               and Stumpf, Gregory J and Smith, Travis M",
  journal   = "Bull. Am. Meteorol. Soc.",
  publisher = "American Meteorological Society",
  volume    =  99,
  number    =  10,
  pages     = "2025--2043",
  month     =  oct,
  year      =  2018,
  url       = "https://doi.org/10.1175/BAMS-D-16-0100.1",
  doi       = "10.1175/BAMS-D-16-0100.1",
  issn      = "0003-0007"
}

@ARTICLE{Roulston2002-eq,
  title     = "Evaluating Probabilistic Forecasts Using Information Theory",
  author    = "Roulston, Mark S and Smith, Leonard A",
  journal   = "Mon. Weather Rev.",
  publisher = "American Meteorological Society",
  volume    =  130,
  number    =  6,
  pages     = "1653--1660",
  month     =  jun,
  year      =  2002,
  url       = "https://doi.org/10.1175/1520-0493(2002)130<1653:EPFUIT>2.0.CO;2",
  doi       = "10.1175/1520-0493(2002)130<1653:EPFUIT>2.0.CO;2",
  issn      = "0027-0644"
}

@ARTICLE{Shannon1948-nc,
  title     = "A mathematical theory of communication",
  author    = "Shannon, C E",
  journal   = "Bell Syst. Tech. J.",
  publisher = "Institute of Electrical and Electronics Engineers (IEEE)",
  volume    =  27,
  number    =  3,
  pages     = "379--423",
  month     =  jul,
  year      =  1948,
  url       = "http://ieeexplore.ieee.org/lpdocs/epic03/wrapper.htm?arnumber=6773024",
  doi       = "10.1002/j.1538-7305.1948.tb01338.x",
  issn      = "0005-8580"
}

@INCOLLECTION{Smets1990-nf,
  title     = "Constructing the pignistic probability function in a context of
               uncertainty",
  author    = "Smets, Philippe",
  booktitle = "Uncertainty in Artificial Intelligence",
  publisher = "Elsevier",
  pages     = "29--39",
  series    = "Machine Intelligence and Pattern Recognition",
  year      =  1990,
  url       = "http://dx.doi.org/10.1016/b978-0-444-88738-2.50010-5",
  doi       = "10.1016/b978-0-444-88738-2.50010-5",
  isbn      =  9780444887382,
  issn      = "0923-0459"
}

@ARTICLE{Sudano2015-rk,
  title         = "Pignistic probability transforms for mixes of low- and
                   high-probability events",
  author        = "Sudano, John J",
  journal       = "arXiv [cs.AI]",
  month         =  may,
  year          =  2015,
  url           = "http://arxiv.org/abs/1505.07751",
  archivePrefix = "arXiv",
  primaryClass  = "cs.AI",
  eprint        = "1505.07751",
  doi           = "10.48550/arXiv.1505.07751"
}

@ARTICLE{Taylor2019-is,
  title     = "Preparing for Doris: Exploring public responses to impact-based
               weather warnings in the United Kingdom",
  author    = "Taylor, Andrea L and Kause, Astrid and Summers, Barbara and
               Harrowsmith, Melanie",
  journal   = "Weather Clim. Soc.",
  publisher = "American Meteorological Society",
  volume    =  11,
  number    =  4,
  pages     = "713--729",
  month     =  oct,
  year      =  2019,
  url       = "http://dx.doi.org/10.1175/WCAS-D-18-0132.1",
  doi       = "10.1175/wcas-d-18-0132.1",
  issn      = "1948-8327,1948-8335",
  language  = "en"
}

@ARTICLE{Taylor2024-pz,
  title     = "The effect of likelihood and impact information on public
               response to severe weather warnings",
  author    = "Taylor, Andrea and Summers, Barbara and Domingos, Samuel and
               Garrett, Natalie and Yeomans, Sophie",
  journal   = "Risk Anal.",
  publisher = "John Wiley \& Sons, Ltd",
  volume    =  44,
  number    =  5,
  pages     = "1237--1253",
  month     =  may,
  year      =  2024,
  url       = "http://dx.doi.org/10.1111/risa.14222",
  doi       = "10.1111/risa.14222",
  pmid      =  37743536,
  issn      = "1539-6924,0272-4332",
  language  = "en"
}

@ARTICLE{Todter2012-ou,
  title     = "Generalization of the Ignorance Score: Continuous Ranked Version
               and Its Decomposition",
  author    = "Tödter, Julian and Ahrens, Bodo",
  journal   = "Mon. Weather Rev.",
  publisher = "American Meteorological Society",
  volume    =  140,
  number    =  6,
  pages     = "2005--2017",
  month     =  jan,
  year      =  2012,
  url       = "https://doi.org/10.1175/MWR-D-11-00266.1",
  doi       = "10.1175/MWR-D-11-00266.1",
  issn      = "0027-0644"
}

@ARTICLE{Trujillo-Falcon2022-du,
  title     = "Creating a Communication Framework for {FACETs}: How
               Probabilistic Hazard Information Affected Warning Operations in
               {NOAA’s} Hazardous Weather Testbed",
  author    = "Trujillo-Falcón, Joseph E and Reedy, Justin and Klockow-McClain,
               Kimberly E and Berry, Kodi L and Stumpf, Gregory J and Bates,
               Alyssa V and LaDue, James G",
  journal   = "Weather, Climate, and Society",
  publisher = "American Meteorological Society",
  volume    =  14,
  number    =  3,
  pages     = "881--892",
  month     =  jul,
  year      =  2022,
  url       = "https://journals.ametsoc.org/view/journals/wcas/14/3/WCAS-D-21-0136.1.xml",
  doi       = "10.1175/WCAS-D-21-0136.1",
  issn      = "1948-8327,1948-8335",
  language  = "en"
}

@ARTICLE{Van-Schaeybroeck2016-sl,
  title     = "A probabilistic approach to forecast the uncertainty with
               ensemble spread",
  author    = "Van Schaeybroeck, Bert and Vannitsem, Stéphane",
  journal   = "Mon. Weather Rev.",
  publisher = "American Meteorological Society",
  volume    =  144,
  number    =  1,
  pages     = "451--468",
  month     =  jan,
  year      =  2016,
  url       = "http://journals.ametsoc.org/doi/abs/10.1175/MWR-D-14-00312.1",
  eprint    = "http://dx.doi.org/10.1175/MWR-D-14-00312.1",
  doi       = "10.1175/mwr-d-14-00312.1",
  issn      = "0027-0644,1520-0493",
  language  = "en"
}

@ARTICLE{Walley1982-pm,
  title     = "Towards a frequentist theory of upper and lower probability",
  author    = "Walley, Peter and Fine, Terrence L",
  journal   = "Ann. Stat.",
  publisher = "Institute of Mathematical Statistics",
  volume    =  10,
  number    =  3,
  pages     = "741--761",
  month     =  sep,
  year      =  1982,
  url       = "https://projecteuclid.org/journals/annals-of-statistics/volume-10/issue-3/Towards-a-Frequentist-Theory-of-Upper-and-Lower-Probability/10.1214/aos/1176345868.short",
  doi       = "10.1214/aos/1176345868",
  issn      = "0090-5364,2168-8966"
}

@ARTICLE{Weijs2010-hg,
  title     = "{Kullback–Leibler} Divergence as a Forecast Skill Score with
               Classic {Reliability–Resolution–Uncertainty} Decomposition",
  author    = "Weijs, Steven V and van Nooijen, Ronald and van de Giesen, Nick",
  journal   = "Mon. Weather Rev.",
  publisher = "American Meteorological Society",
  volume    =  138,
  number    =  9,
  pages     = "3387--3399",
  month     =  apr,
  year      =  2010,
  url       = "https://doi.org/10.1175/2010MWR3229.1",
  doi       = "10.1175/2010MWR3229.1",
  issn      = "0027-0644"
}

@ARTICLE{Weijs2011-cf,
  title     = "Accounting for Observational Uncertainty in Forecast
               Verification: An Information-Theoretical View on Forecasts,
               Observations, and Truth",
  author    = "Weijs, Steven V and van de Giesen, Nick",
  journal   = "Mon. Weather Rev.",
  publisher = "American Meteorological Society",
  volume    =  139,
  number    =  7,
  pages     = "2156--2162",
  month     =  mar,
  year      =  2011,
  url       = "https://doi.org/10.1175/2011MWR3573.1",
  doi       = "10.1175/2011MWR3573.1",
  issn      = "0027-0644"
}

@BOOK{Wilks2011-vw,
  title     = "Statistical Methods in the Atmospheric Sciences",
  author    = "Wilks, Daniel S",
  publisher = "Academic Press",
  month     =  may,
  year      =  2011,
  url       = "https://play.google.com/store/books/details?id=IJuCVtQ0ySIC",
  isbn      =  9780123850225,
  language  = "en"
}

@ARTICLE{Williams2014-jb,
  title     = "A comparison of ensemble post‐processing methods for extreme
               events",
  author    = "Williams, R M and Ferro, C A T and Kwasniok, F",
  journal   = "Quart. J. Roy. Meteor. Soc.",
  publisher = "Wiley",
  volume    =  140,
  number    =  680,
  pages     = "1112--1120",
  month     =  apr,
  year      =  2014,
  url       = "https://rmets.onlinelibrary.wiley.com/doi/10.1002/qj.2198",
  doi       = "10.1002/qj.2198",
  issn      = "0035-9009,1477-870X",
  language  = "en"
}

@BOOK{Yager2005-at,
  title     = "Classic works of the Dempster-Shafer theory of belief functions",
  editor    = "Yager, Ronald R and Liu, Liping",
  publisher = "Springer",
  address   = "Berlin, Germany",
  edition   =  2008,
  series    = "Studies in Fuzziness and Soft Computing",
  month     =  may,
  year      =  2005,
  url       = "https://link.springer.com/book/10.1007/978-3-540-44792-4",
  doi       = "10.1007/978-3-540-44792-4",
  isbn      = "9783540253815,9783540447924",
  language  = "en"
}

@ARTICLE{Zadeh1978-je,
  title     = "Fuzzy sets as a basis for a theory of possibility",
  author    = "Zadeh, L A",
  journal   = "Fuzzy Sets and Systems",
  publisher = "Elsevier",
  volume    =  1,
  number    =  1,
  pages     = "3--28",
  month     =  jan,
  year      =  1978,
  url       = "https://www.sciencedirect.com/science/article/pii/0165011478900295",
  doi       = "10.1016/0165-0114(78)90029-5",
  issn      = "0165-0114"
}

@INCOLLECTION{Chevrie1998-go,
  title     = "Fuzzy Logic",
  author    = "Chevrie, F and Guély, F",
  publisher = "Groupe Schneider",
  address   = "Grenoble",
  month     =  dec,
  year      =  1998
}


\end{document}